\documentclass[a4paper,11pt]{article}
\usepackage[T1]{fontenc}
\usepackage{lmodern}
\usepackage[utf8]{inputenc}
\usepackage{authblk}
\usepackage{setspace}
\usepackage[margin=1.08in,includefoot]{geometry}
\usepackage{graphicx}
\usepackage{subcaption}
\usepackage{amsmath,amssymb,amsfonts}
\usepackage{float}
\usepackage{booktabs}
\usepackage{makecell}
\usepackage[table]{xcolor}
\usepackage{physics}
\usepackage{soul}
\usepackage{tikz}
\usepackage{quantikz}
\usepackage[protrusion=false,expansion=false]{microtype}
\usepackage[hidelinks]{hyperref}

\usepackage[style=phys,citestyle=numeric-comp,sorting=none,backend=biber,eprint=true,url=true]{biblatex}
\newcommand{\mainthree}[1]{#1}

\hypersetup{pdftitle={How Far Can You Do Nothing On a Quantum Computer?},pdfauthor={Nitay Mayo; Tal Mor; Aryeh Lev Zabokritskiy (Yohananov)}}
\newcommand{\SupplementMap}[3]{%
  \begin{figure}[p]
    \centering
    \includegraphics[page=#1,width=\linewidth,trim=79bp 0bp 79bp 0bp,clip]{figures.pdf}
    \caption{#2}
    \label{#3}
  \end{figure}
  \clearpage
}

\title{How Far Can You Do Nothing On a Quantum Computer?}

\author{Nitay Mayo$^1$\thanks{Corresponding author: \href{mailto:nitay.mayo@campus.technion.ac.il}{nitay.mayo@campus.technion.ac.il}.}, Tal Mor$^{1,2}$ and Aryeh Lev Zabokritskiy (Yohananov)$^3$}
\affil{$^1$Department of Computer Science, Technion University, Haifa, Israel.}
\affil{$^2$The Helen Diller Quantum Center}
\affil{$^3$Department of Computer Science, MIGAL – Galilee Research
Institute/Tel-Hai University of Kiryat Shmona and the Galilee, Kiryat
Shmona, Israel}
\date{}

\begin{document}
\maketitle
\begin{abstract}
We present a route-resolved comparative assessment of Rigetti's Cepheus-1-108Q and IBM Heron-r2 processors using the established \textit{do-nothing} state-transfer protocol. Rather than proposing a new protocol, we use this deterministic, low-complexity task as a high-resolution spatial probe. For each evaluated initial qubit, we report two complementary quantities: the largest tested radius within which every evaluated shortest route satisfies the operational success rule, and the longest successful route identified within the evaluated route family.
To address the question ``\textit{How far can you do-nothing on a quantum computer?}'', we prepare a quantum state, route it outward using SWAP gates, apply the inverse preparation at the destination, and route it back for measurement. We compare the measured probability of returning $0$ with the empirical cutoff $2/3$. For this circuit ordering, that comparison does not certify above-classical unknown-state transfer.
While this trivial state-transfer protocol serves as the most intuitive baseline, actively preserving a quantum state across a physical lattice proves to be a non-trivial task that exposes the information to cumulative relaxation, dephasing, and environmental cross-talk.
In the highlighted IBM QPU case, we identify an isotropic radius of 10 and a successful path of swap distance 27, whereas the highlighted Rigetti Cepheus case exhibits an isotropic radius of 1 but selected above-threshold routes reaching swap distance 8. These results reveal a sharp distinction between uniform spatial reliability and best-route performance. The presented quantities are empirical and conditional on the evaluated route families, finite-shot decision rule, calibration state, and execution time; they are not architecture-wide constants.
\end{abstract}
\newpage
\tableofcontents

\section{Introduction}
Recent advancements in quantum computing have transitioned the field from theoretical exploration to the demonstration of significant computational advantages. Notable milestones include the first claims of quantum supremacy via random-circuit sampling~\cite{Arute_2019}
 and the demonstration of quantum utility beyond brute-force classical simulation using error-mitigated superconducting circuits~\cite{kim2023evidence}. These achievements underscore the rapid scaling of Quantum Processing Units (QPUs) and the increasing complexity of the tasks they can perform. Despite these monumental achievements, the field remains firmly within the \textit{Noisy Intermediate-Scale Quantum} (NISQ) era~\cite{Preskill_2018}, where unmitigated environmental noise, topological limitations, and gate imperfections heavily constrain algorithmic execution.
As benchmarking methods become more intricate to keep pace with these developments, understanding the underlying information in their results is increasingly difficult~\cite{Proctor_2021}. 
We believe that an honest, simple assessment of hardware is essential for the practical development of capable quantum processors and algorithms.

For this task, we employ the \textit{do-nothing} protocol~\cite{meirom2025} as a routing benchmark: we prepare a quantum state, route it outward via SWAP gates, apply the inverse preparation, and route it back to measure the probability of outcome $0$.
While typical quantum circuits are designed to execute complex, nontrivial computations, evaluating the spatial limitations of a processor requires a different approach. Consider a practical algorithmic scenario: a qubit initially at location $A$ is required to act as a control qubit for an operation on a register situated near a distant location $B$. Then this qubit must be returned to its original position, unmodified or with as minimal error as possible. This necessitates routing the state to location $B$, executing the control operation, and routing it back to location $A$. Routing errors can affect such an operation. The almost empty circuit studied here probes route-dependent outcomes, although its final return score alone does not establish preservation of an arbitrary unknown state for that operation. 

Our results show that although ``doing nothing'' is a simple baseline, its physical execution is far from trivial, as it encapsulates the cumulative degrading effects of $T_1$ relaxation, $T_2$ dephasing, and environmental crosstalk~\cite{Sarovar_2020, Ding_2020}. The resulting return scores provide a simple route-resolved diagnostic of this combined noise.
Previous works in the '\textit{Benchmarking via Protocols}' series introduced a comprehensive evaluation framework and a suite of protocol-based tests~\cite{mayo2026_1, mayo2026_2, meirom2025}. Those initial studies established the theoretical foundation and articulated a selection logic for linear paths and constrained sub-chips.
Using this framework and prior IBM evaluations to inform the Rigetti experimental design, we systematically extend the route selection across the processor. We retain the numerical cutoff $2/3$, motivated by the classical average-fidelity bound for unknown single-qubit state transfer~\cite{Massar_Popescu_1995}, but use it here only as an empirical classification rule for the measured return score.

\subsection{\texorpdfstring{Relation to Existing Benchmarks and Scope of the Contribution}{Relation to Existing Benchmarks and Scope of the Contribution}}
Cross-platform and full-stack quantum-computer benchmarks already compare devices, qubit subsets, and complete execution pipelines. However, it is important to clarify how our specific spatial focus differs from established metrics. For instance, holistic metrics like Quantum Volume~\cite{pelofske_qv_2022} yield a single, chip-wide score, while Randomized Benchmarking focuses on averaged, gate-level error rates. Neither approach isolates the specific spatial routing constraints, directional dependencies, or topological overheads encountered during unmitigated state transfer. Other representative examples include application-oriented full-stack suites~\cite{lubinski_app_bench_2023,hines_fullstack_2024}, the SupermarQ suite~\cite{tomesh_supermarq_2022}, and single-qubit comparison protocols~\cite{suau_single_qubit_2023}. While recent layer-fidelity work selects and monitors long physical qubit chains~\cite{lozano_layer_fidelity_2026} and QKNOB isolates connectivity and routing overhead in circuit transformation~\cite{li_qknob_2025}, our approach maps the spatial distribution of above-cutoff return scores for the tested protocol.

The broader idea of using communication protocols as hardware benchmarks emerged around 2017–2019, explored independently by Zhukov \textit{et al.}—who applied an entropic threshold to the superdense coding protocol (and more) on IBM processors and simulators~\cite{zhukov_protocol_benchmark_2019}—and within Tal Mor's research group, who applied fidelity and entanglement thresholds to teleportation and entanglement swapping protocols, respectively. These thresholds motivated the terminology of quantum regimes, quantum sub-chips, and quantum protocol vectors in that framework. The star-graph quantities used here classify empirical return scores and do not certify quantumness.

Specifically, early hardware evaluations were conducted within Mor's group by Rotem Liss and Chen Mechel. Mor presented these results at several universities and workshops under titles such as ``Quantum computing—is the future here?'' and ``How many qubits are there, in my $n$-qubit quantum computer?''. In particular, presentations were given at the Qubit workshop celebrating C. H. Bennett and G. Brassard's Wolf Prize for Physics (Technion, Haifa, Israel, June 2018), as well as at CWI Amsterdam, in Aachen, and in Vienna during the summer of 2018. This continuous effort led to the exploration of extended ideas, culminating in the formal \textit{Benchmarking via Protocols} framework. This framework generalized the approach to multi-qubit protocols and introduced evaluations via superdense coding, Bell state transfer, the \textit{transmit} protocol, and the \textit{do-nothing} baseline~\cite{meirom2025, mayo2026_1, mayo2026_2}.

Regarding evaluations on other hardware, recent studies have also successfully run communication protocols, such as teleportation, to compare IBM and Rigetti processors~\cite{marquez_teleportation_benchmark_2025}. Building upon these foundations, our progression leverages the strict \textit{do-nothing} baseline to systematically map spatial capabilities, representing the first application of this specific evaluation on Rigetti's architecture alongside a comparison to IBM's different qubit connectivity.

Accordingly, the contribution claimed here is neither the first cross-platform benchmark nor the first use of a communication protocol, routing, or spatially selected qubits for hardware assessment. The contribution is the route-resolved use of a fixed state-transfer task to separate two operationally different spatial properties: uniform reliability over every evaluated shortest route inside a Sub-Star chip — defined as a specific subset of qubits centered around an initial qubit, relative to which all spatial metrics are measured — and exceptional long-route performance within the tested route family. This distinction, together with explicit maps of the contributing qubits and couplers, is the central novelty evaluated in this work.

\section{Methodology}\label{sec:method}
\subsection{The ``Do-Nothing'' protocol and Mathematical Framework}
As said, the ``\textit{do-nothing}'' protocol is a fundamental benchmarking tool we use to characterize the spatial noise profile of a QPU. Rather than merely isolating intrinsic idle-qubit decoherence, the measured degradation in this protocol captures the cumulative impact of SWAP-gate errors, routing-dependent noise, environmental crosstalk, and state preparation and measurement (SPAM) effects. Originally, this protocol was proposed by Meirom, Mor, and Weinstein~\cite{meirom2025}. The protocol's utility was subsequently used to benchmark two distinct IBM quantum processors~\cite{mayo2026_1} and further generalized to contrast IBM architectures with AQT's ion-trap system \cite{mayo2026_2}. However, for the do-nothing protocol, these earlier evaluations were limited to coarse spatial segmentation of the tested QPUs. In the present work, we remove these topological constraints, deploying the \textit{do-nothing} protocol to evaluate unconstrained sections on the processor grid.
As illustrated in Figure~\ref{fig:do_nothing_protocol}, the work qubit is initialized in $\ket{0}$ and a unitary $U_{rand}$ prepares $\ket{\psi}=U_{rand}\ket{0}$. A sequence of $L$ SWAP gates routes the work state to a destination qubit. We apply $U_{rand}^{\dagger}$ there and route the state back to the initial qubit. If $\rho'$ is its final state, the measured return probability is $\mathcal{F}=\bra{0}\rho'\ket{0}$, estimated by $\widehat{\mathcal{F}}=n_0/512$, where $n_0$ counts work-qubit outcomes equal to zero. This is the output-state fidelity with $\ket{0}$; it is not in general the fidelity of the outward transferred state with the unknown input $\ket{\psi}$. We use ``return score'' for its measured estimate below.

The distance $L$, namely ``swap distance'', is the number of SWAP gates applied in order to transfer the state to the furthest point in the circuit. For example, in Figure~\ref{fig:do_nothing_protocol}, the swap distance is three.

The classical bound $2/3$ concerns average fidelity for estimating and reproducing an unknown, uniformly distributed pure qubit from one copy~\cite{Massar_Popescu_1995}. It does not generally bound the return score of this circuit, because $U_{rand}^{\dagger}$ precedes the noisy return route. For example, if each transfer direction replaces its input by $\ket{0}\!\bra{0}$, the final return probability is $1$ for every $U_{rand}$, although the outward transfer preserves no input information. We therefore use $2/3$ solely as an empirical cutoff, without claiming certification of quantum advantage or above-classical unknown-state transfer.

A tested route is classified using its return score and the prefix rule defined below. Throughout the paper, ``successful'' and ``above-threshold'' refer only to that empirical rule.

\begin{figure*}[!t]
    \centering
    \begin{quantikz}[row sep={0.8cm,between origins}]
    \lstick{$q_0$} & \gate{U_{rand}} & \swap{1}  & \qw      & \qw      & \qw                      & \qw      & \qw       & \swap{1} & \meter{} \vcw{4} & \qw \\
    \lstick{$q_1$} & \qw             & \targX{} & \swap{1} & \qw      & \qw                      & \qw      & \swap{1}  & \targX{} & \qw & \qw\\
    \lstick{$q_2$} & \qw             & \qw      & \targX{} & \swap{1} & \qw                      & \swap{1} & \targX{} & \qw       & \qw & \qw\\
    \lstick{$q_3$} & \qw             & \qw      & \qw       & \targX{} & \gate{U_{rand}^\dagger} & \targX{} & \qw      & \qw       & \qw & \qw \\
    \lstick{c}     & \cw             & \cw      & \cw       & \cw       & \cw                     & \cw      & \cw       & \cw       & \cw & \cw
    \end{quantikz}
\caption{The \textit{do-nothing} protocol at swap distance $L=3$. The state is prepared at $q_0$, routed to $q_3$, acted on by $U_{rand}^{\dagger}$, and routed back to $q_0$ for measurement.}
\label{fig:do_nothing_protocol}
\end{figure*}

\subsection{Experimental Design}\label{sec:experimental_design}
In this work, we evaluate the Rigetti Cepheus-1-108Q and the IBM quantum processors, specifically focusing on the IBM Fez QPU. To ensure a clear comparative analysis, the main text prioritizes the detailed spatial evaluations of the Rigetti architecture alongside a selected initial qubit from the IBM Fez chip, while additional IBM data is provided in the supplementary materials. To accurately answer the core research question across the tested QPUs, we implemented an experimental design directly informed by the conclusions of prior \textit{Benchmarking via Protocols} studies~\cite{meirom2025, mayo2026_1, mayo2026_2}. Those earlier evaluations demonstrated that accumulated SWAP-gate errors rapidly degrade state fidelity. Consequently, our procedure was structured as follows:

\begin{enumerate}
    \item \textbf{Initial Qubits Selection:} We selected the \textit{initial qubits} sub-set - these qubits are chosen on each chip and serve as the initial node from which we execute the protocol. This spatial distribution was deliberately chosen to capture potential performance variances between different sections of the chip. Specifically, to evaluate edge-case topological constraints, we selected four representative qubits on the Rigetti chip (0, 40, 67 and 107): two located centrally within the QPU and two at opposite peripheral corners. Leveraging a larger experimental budget for the IBM hardware, we expanded this sampling logic to include eight \textit{initial qubits}, comprising four central qubits and four at the corners.

    \item \textbf{Executing the Protocol:} From each initial qubit, the \textit{do-nothing} protocol was executed, targeting other accessible qubits on the processor in all directions. In all instances, state transfer was restricted exclusively to the shortest topological paths. \mainthree{Each circuit was executed with 512 shots. A route circuit passes the pre-specified endpoint cutoff when $\widehat{\mathcal{F}} > 2/3$; a \emph{successful path} must additionally satisfy the prefix rule below. This is an empirical classification, not a hypothesis test. Near the threshold, the binomial standard error is approximately $\sqrt{(2/3)(1/3)/512}\simeq 0.021$, so routes separated from the threshold by only a few percentage points are boundary-sensitive. All radii and path lengths reported below use the same empirical rule, and no confidence-corrected reclassification is implied.} The same numerical value appears in state-transfer experiments such as quantum teleportation~\cite{Pfaff_2014}, but their classical-bound interpretation does not transfer automatically to the circuit used here. The retained data contain one count record per route, but do not specify the preparation-unitary distribution, the number of independent preparations contributing to each record, or any averaging across preparations. These scores are therefore empirical summaries of the recorded experiments, without a verified input-ensemble average. The quoted binomial uncertainty describes shot sampling only; variation across preparation states cannot be quantified from the retained records.

    \item \textbf{Path Filtering and Success Criteria:} We enforce a prefix-closed empirical criterion. A route is a ``successful path'' only if the separately executed circuit for that route and the circuits for every prefix from the same initial qubit each have return score $\widehat{\mathcal{F}}>2/3$. These are final measurements at the initial qubit after the corresponding round trips, not measurements taken along one execution. In the staged Rigetti search, routes failing the rule at shorter distances were not extended.

\end{enumerate}

Our procedure yields two distinct primary metrics, each answering a different question regarding the spatial capabilities of the hardware. We allow a radius of zero and write $\mathbb{N}_0=\{0,1,2,\ldots\}$:\begin{itemize}
    \item \textbf{Maximal Isotropic Length:} This metric defines the region around an initial qubit $q_0$ in which every evaluated shortest route passes the empirical rule. Formally, for every two qubits $q_0$ and $q_t$, let $d(q_0,q_t)$ denote the swap distance between them and $P(q_0,q_t)$ be the set of all evaluated shortest paths connecting them. Then for each initial qubit $q_0$, the \textit{Maximal Isotropic sub-star} radius $r_{iso}$ is defined as:
    \begin{equation}
        \begin{split}
        r_{iso} = \max \{&r\in \mathbb{N}_0 \ | \ \forall q_t\ \text{such that}\\
        &\ d(q_0,q_t) \le r, \ \forall p\in P(q_0,q_t),\\
        &\ p\ \text{is successful} \}
        \end{split}
    \end{equation}
    
    For every target within distance $r$, \textit{every} tested shortest route must pass the empirical rule. The metric may inform noise-aware placement and routing~\cite{murali2019noiseadaptivecompilermappingsnoisy}, but tests from a single initial qubit do not establish arbitrary pairwise transfers or guarantee algorithmic reliability within the region. The initial qubit and all qubits within $r_{iso}$ form the \textit{Maximal Isotropic Sub-star}.

    \item \textbf{Maximal Successful Path:} Distinct from the isotropic region, this metric identifies the existence of long individual routes passing the empirical rule. It represents the longest single topological trajectory from the tested routes set originating from the initial qubit that fully satisfies the criteria for a successful path. With the same notation as the former metric, for each $q_0$ the \textit{Maximal Successful Length} $r_{max}$ is defined as:
    \begin{equation}
        \begin{split}
        r_{max} = \max \{&r \in \mathbb{N}_0 \ | \exists q_t\ \text{such that}\\
        &\ r = d(q_0,q_t) \ \text{and}\ \exists p \in P(q_0, q_t),\\
        &\ p\ \text{is successful} \}
        \end{split}
    \end{equation}
    This metric records the longest identified route satisfying the empirical prefix rule, even if neighboring routes fail.

\end{itemize}

The staged strategy should therefore be interpreted as a cost-saving search heuristic motivated by the observed overall degradation trend, not as a proof of pathwise monotonicity. In particular, the Rigetti maximal successful length is the longest route identified by the staged search under the stated execution conditions; it is not asserted to be the architecture-wide longest route that could exist under an exhaustive search or a different calibration state. More generally, both spatial metrics are conditional on the explicitly evaluated path family.

In each violin plot below, the grey distribution, colored points and yellow maximum use the same set of recorded endpoint scores for the four initial qubits listed in that panel, including routes that fail the prefix rule. Each route has equal weight. The violins use Gaussian kernel-density estimates with Scott's bandwidth; their widths do not encode route counts. Small horizontal offsets separate overlapping points without changing their integer swap distances.

\section{Results of Rigetti's QPU}\label{sec:comparison_section}

\begin{figure*}[!t]
    \centering
    \includegraphics[page=1,trim=88bp 38bp 88bp 38bp,clip,width=\linewidth]{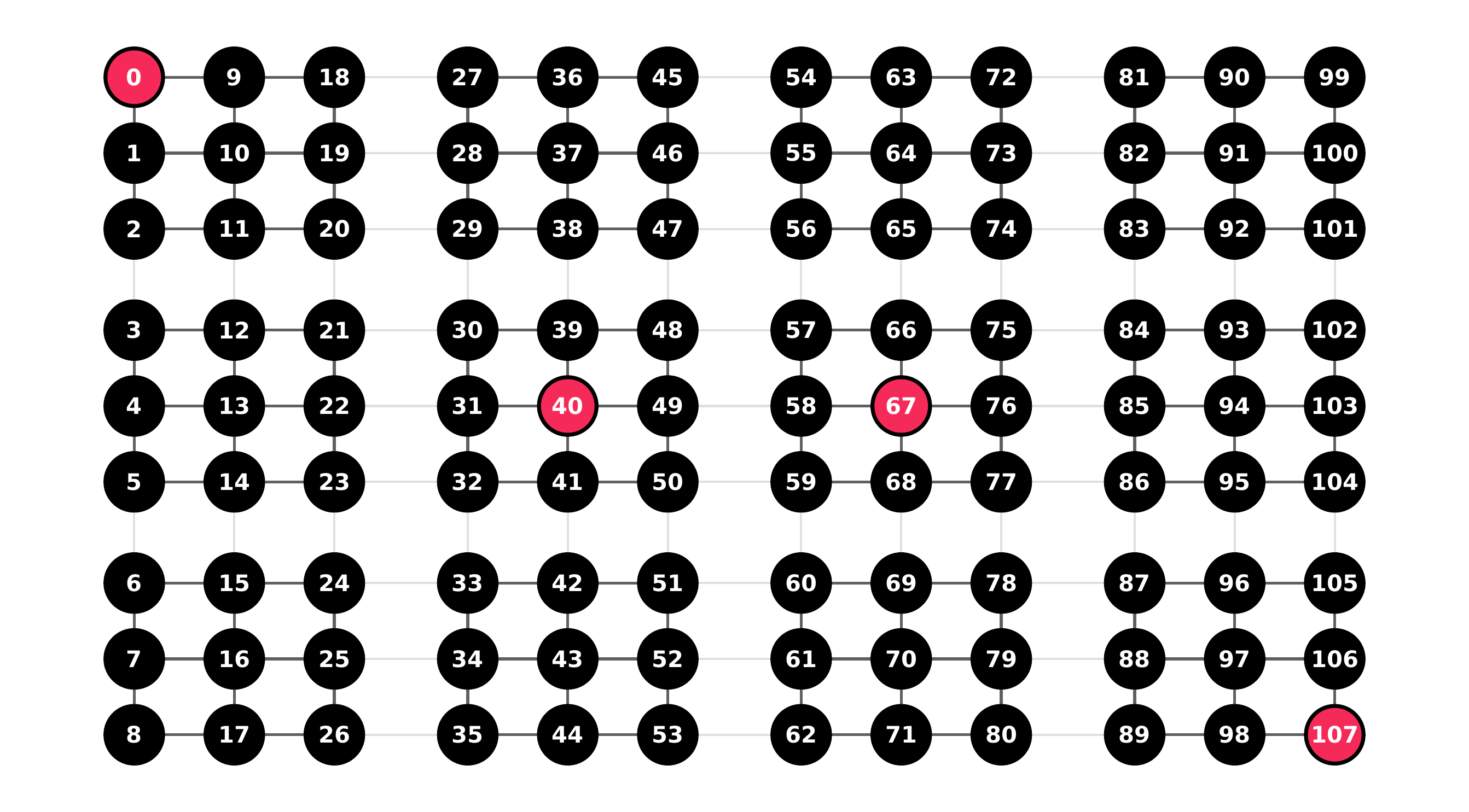}
    \caption{Cepheus-1-108Q connectivity map. The chosen initial qubit set is colored red. In this graph, each node represents a qubit, and each edge is a coupler, intra-chiplet (dark grey) or IMC (light grey).}
    \label{fig:cepheus_map}
\end{figure*}

At the time of our experiments, the most recent iteration of Rigetti Computing's Cepheus-1-108Q represents an evolution in QPU design by employing a multi-chip architecture that tiles twelve (a three-by-four grid) distinct 9-qubit chiplets (Fig~\ref{fig:cepheus_map}), culminating in a 108-qubit system~\cite{rigetti_cepheus_manual}.
The primary strength of this modular, chiplet-based approach lies in its inherent scalability. Fabricating massive quantum processors is notoriously challenging due to strict physical constraints, intense calibration demands, and more engineering challenges. By segmenting the processor into smaller 9-qubit chiplets that can be fabricated, tested, and optimized independently~\cite{norris2025performancecharacterizationmultimodulequantum}, Rigetti has established a pragmatic manufacturing approach for future systems. However, this design inherently introduces complex physical limitations on the connection couplers between chiplets. To create the 108-qubit grid, the system relies on inter-module Couplers (IMCs) that bridge between the chiplets. These IMCs, colored light grey in Figure~\ref{fig:cepheus_map}, may act differently on the quantum information than the intra-chiplet couplers,
potentially exhibiting different noise profiles~\cite{zhao2026longrangetunablecouplermodular}, susceptibility to environmental cross-talk~\cite{gold2021entanglementseparatesilicondies}, and varying gate fidelities compared to the inner connections within a single chiplet.
Consequently, running the \textit{do-nothing} circuit across IMC boundaries probes routes spanning chiplets, without separately identifying the contribution of an individual coupler.

\begin{figure*}[p]
    \centering
    \begin{subfigure}{\linewidth}
        \centering
        \includegraphics[page=2,trim=88bp 38bp 88bp 38bp,clip,width=\linewidth]{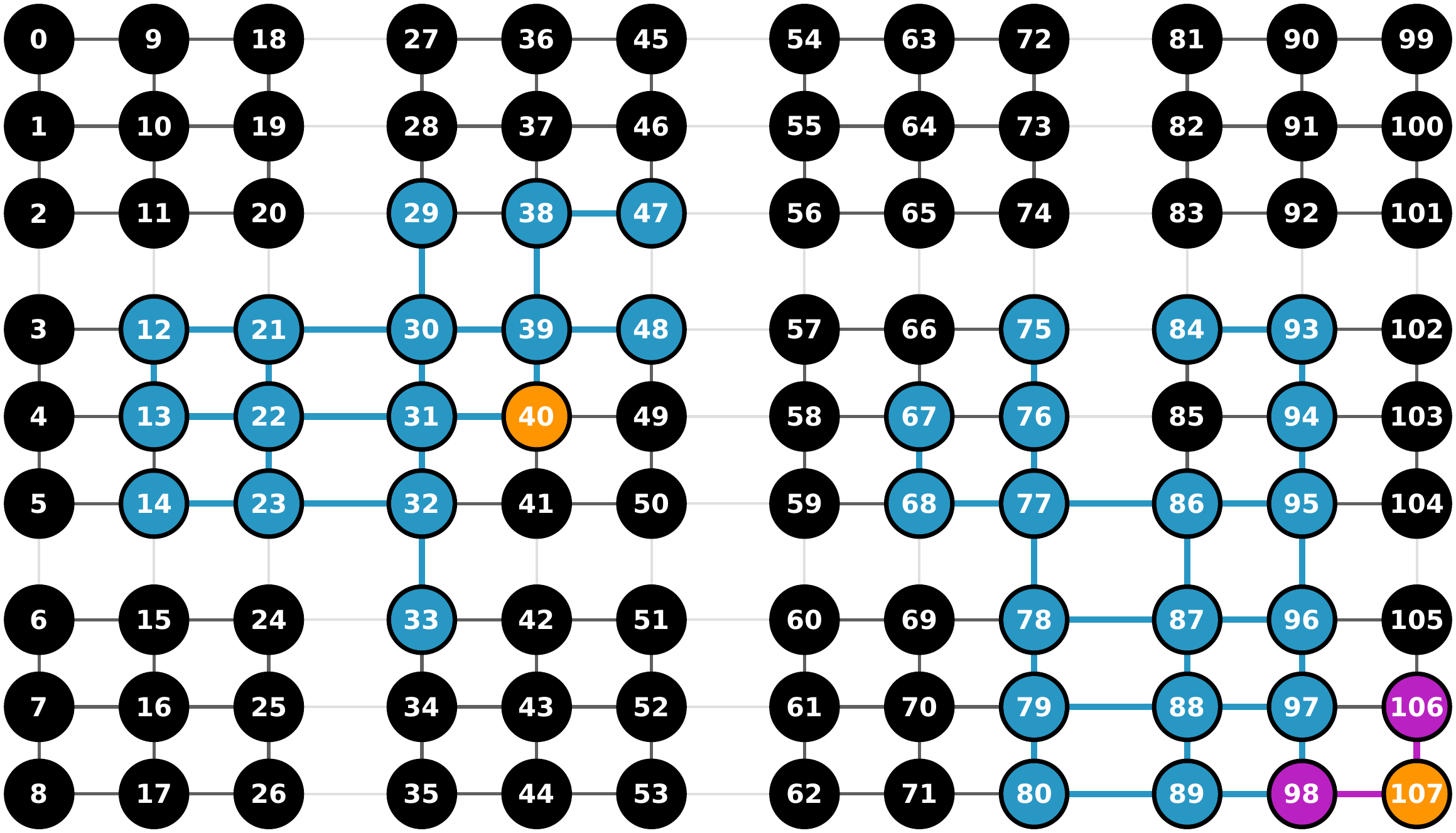}
        \caption{Initial qubits 40 and 107.}
    \end{subfigure}
    \par\smallskip
    \begin{subfigure}{\linewidth}
        \centering
        \includegraphics[page=3,trim=88bp 38bp 88bp 38bp,clip,width=\linewidth]{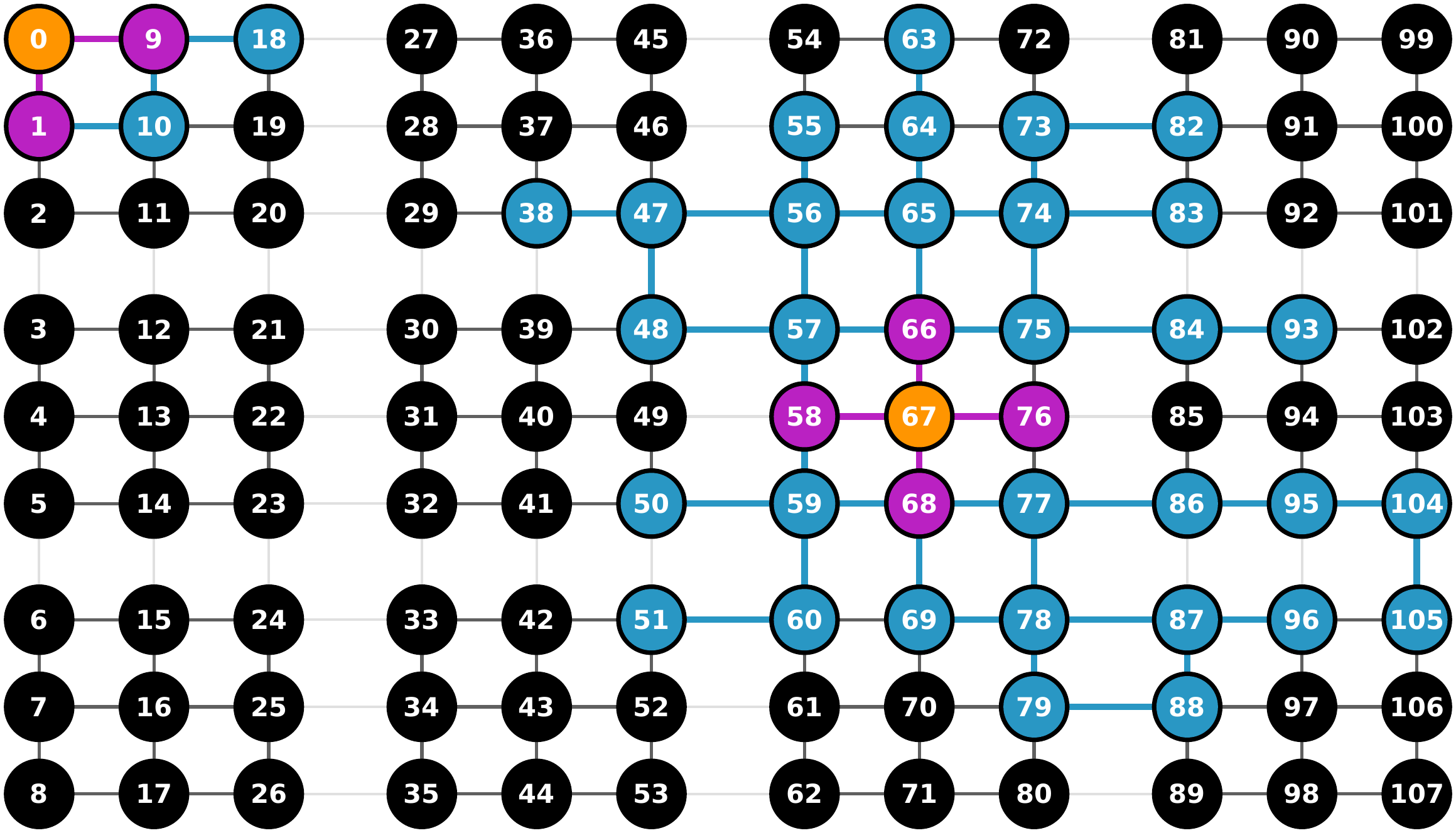}
        \caption{Initial qubits 0 and 67.}
        \label{fig:cepheus_spider_charts_0_67}
    \end{subfigure}
    \caption{Overlay of successful paths on the Cepheus grid. The initial qubits are highlighted in orange. Light blue nodes and edges indicate components contained within at least one successful path from an initial qubit to a target qubit. Magenta nodes represent the \textit{Maximal Isotropic Sub-star}.}
    \label{fig:cepheus_spider_charts}
\end{figure*}

To evaluate the Cepheus-1-108Q architecture, we designated four \textit{initial qubits} (highlighted in red, Figure~\ref{fig:cepheus_map}) from which the \textit{do-nothing} protocol was routed only along the shortest topological paths between them and other qubits.
To map the evaluated routes in Figure~\ref{fig:cepheus_spider_charts}, light blue nodes and edges denote components belonging to at least one successful path. The magenta \textit{Maximal Isotropic Sub-star} requires every evaluated shortest route within its radius to pass. For initial qubit 40, the isotropic radius is zero: the one-step routes to qubits 41 and 49 have return scores $245/512$ and $313/512$, respectively, both below $2/3$. Routes to qubits 31 and 39 pass the cutoff and can belong to the blue successful-route set, but do not create an isotropic radius of one. The isotropic regions are small, and some routes fall below the return-score cutoff even at short distances, including intra-chiplet routes from qubit 0. Selected routes from qubits 107 and 67 nevertheless satisfy the prefix rule through swap distances eight and six, respectively. These observations distinguish uniform performance over tested routes from the performance of selected long routes; they do not independently establish coherent state preservation.

To systematically map these limits while minimizing experimental overhead, we deployed a staged execution strategy to identify the longest successful path for each initial qubit. Initially, we evaluated all paths up to a swap distance of four.
Only those topological routes passing the return-score rule $\widehat{\mathcal{F}} > 2/3$ were extended to all paths of length eight, and finally ten. These specific cutoffs ($L=4$, $L=8$, and $L=10$) were deliberately chosen to balance rigorous spatial mapping with QPU access and budget constraints, and were also informed by the average degradation trend observed in both our preliminary Rigetti evaluations and our previous IBM experiments.
At $L=10$, the return scores of all remaining paths fell below the empirical cutoff. This boundary concluded the experimental process, allowing us to locate the longest viable paths originating from each initial qubit.

\begin{figure*}[t]
    \centering

    \includegraphics[page=4,width=\linewidth]{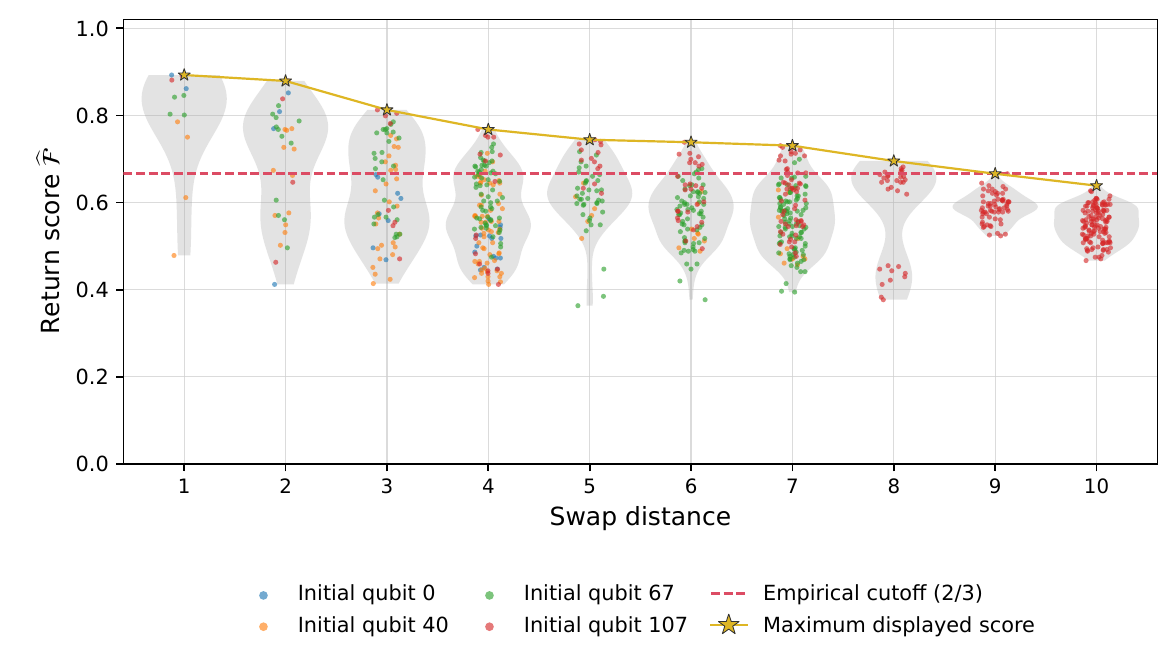}
    
    \caption{Return score as a function of swap distance on the Cepheus-1-108Q chip. Grey violins summarize the recorded route scores for the four initial qubits listed in this panel, while scatter points detail individual route outcomes color-coded by their initial qubit (0, 40, 67, and 107). The red dashed line marks the empirical cutoff ($\mathcal{F} = 2/3$), and the yellow trajectory tracks the largest displayed return score at each swap distance.}
    \label{fig:cepheus_fid_vs_swap_dist_chart}
\end{figure*}

{Figure~\ref{fig:cepheus_fid_vs_swap_dist_chart} shows return-score distributions at each swap distance for the evaluated Cepheus routes. The distributions shift toward lower scores as distance increases, while selected routes retain above-cutoff scores. The yellow curve gives the largest displayed return score at each distance. A point above $2/3$ does not by itself establish a successful route: every prefix must also pass the rule in Section~\ref{sec:experimental_design}. The longest successful routes identified by the staged search reach swap distance eight.}

A route from qubit 67 satisfies the prefix rule through swap distance six, and some shortest routes of length eight from qubit 107 to qubits 67 and 75 also pass. The spread between average and maximum return scores shows that selected routes can perform differently from the route population. Cepheus offers many alternative topological routes, but these data do not isolate a causal effect of connectivity or establish how a compiler would perform on another task.

The evaluated Cepheus routes therefore show limited uniform above-cutoff performance alongside selected longer routes that pass the empirical rule across chiplets.

Having established these spatial routing capabilities on the Rigetti Computing hardware, we turn to our comparative evaluation of the IBM Quantum processor.

\section{Results of IBM's QPU}\label{sec:ibm_results}
\begin{figure*}[htbp]
    \centering
    \includegraphics[page=5,width=\linewidth]{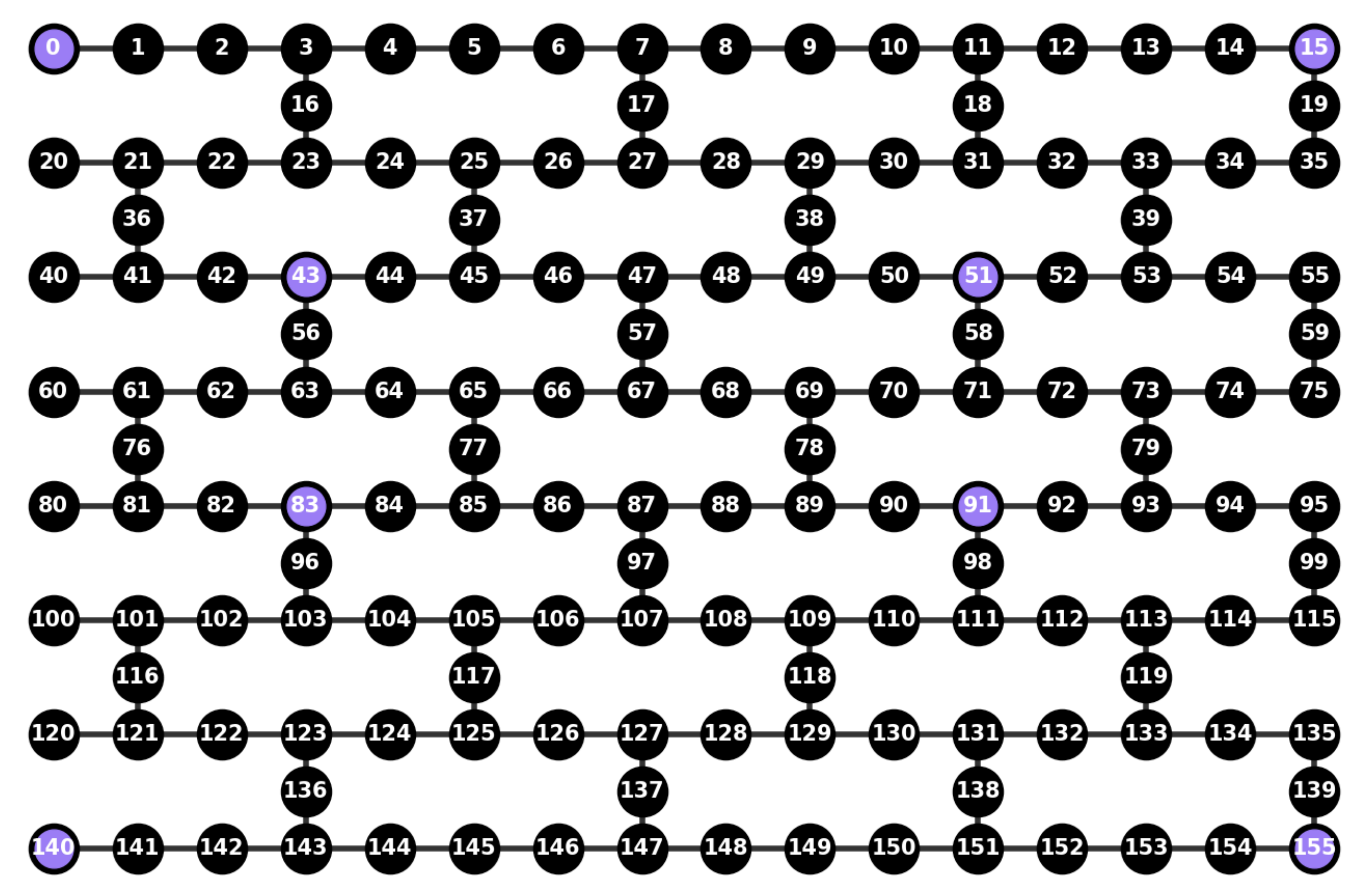}
    \caption{IBM's Fez Heron r2 connectivity map. The chosen initial qubit set is colored purple}
    \label{fig:heron_r2_map}
\end{figure*}

During this research, we tested three QPUs from IBM: Fez, Kingston, and Marrakesh, which were publicly available on the IBM quantum cloud platform at the time of the experiments.
Although the full IBM experiment covered multiple initial qubits on Fez, Kingston, and Marrakesh, the main text displays Fez qubit 140 as a deliberately high-performing illustrative case selected after inspection of the IBM results. It was chosen because it combines a large isotropic region, a visible boundary in the successful-route map, and an exceptionally long successful path; it should not be interpreted as an average or pre-selected representative qubit. Across the evaluated IBM initial qubits, the mean isotropic radius is approximately 7, and the mean maximal successful length is 21.8, whereas qubit 140 attains 10 and 27, respectively. Additional spatial maps and return-score-versus-swap-distance plots are provided in the appendices (Sections~\ref{sec:fez_supp_chart}--\ref{sec:additional_swap_distance}). Consequently, the main IBM--Rigetti comparison below is an illustrative best-case comparison, not an unbiased ranking of the two vendors or architectures.

Figure~\ref{fig:heron_r2_map} presents the connectivity map of the tested IBM QPU.
The qubits marked in purple are the designated \textit{initial qubits} as explained in section~\ref{sec:experimental_design}.
This graph illustrates the connectivity constraints of IBM's QPUs; each qubit can directly interact only with its nearest neighbors, which are at most three due to the heavy-hex topological structure.

\subsection{Results from the Fez QPU}\label{sec:fez_results}

\begin{figure*}[t]
    \centering

    \includegraphics[page=6,width=\linewidth]{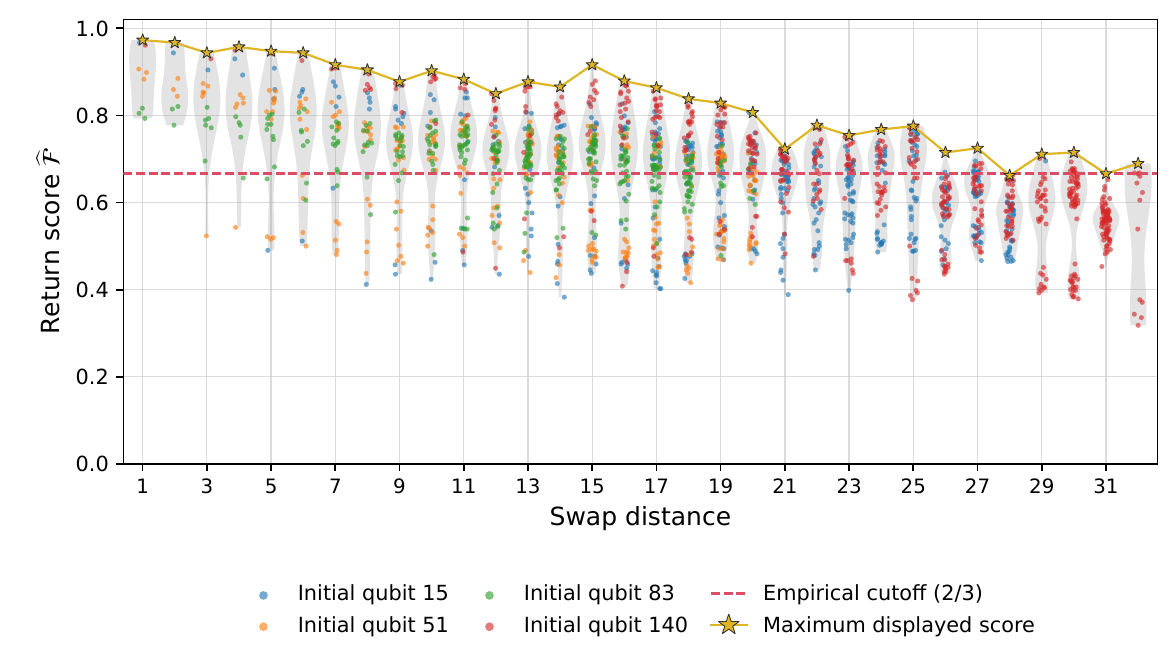}
    \caption{Return score as a function of swap distance on the IBM Fez processor. Grey violins summarize the recorded route scores for the four initial qubits listed in this panel, while scatter points detail individual route outcomes color-coded by four initial qubits (15, 51, 83, and 140). Qubit 140 is highlighted as a high-performing case, rather than as a typical initial qubit. The red dashed line marks the empirical cutoff ($\mathcal{F}=2/3$), and the yellow trajectory tracks the largest displayed return score at each swap distance. Data for the other four initial qubits are provided in Fig.~\ref{fig:fez_fid_vs_swap_dist_chart_2} in Appendix~\ref{sec:fez_supp_chart}.}
    \label{fig:fez_swap_dist}
    
\end{figure*}

As previously established, the experimental process began by executing the \textit{do-nothing} protocol from the eight initial qubits. Crucially, rather than selecting a single route, we evaluated all possible shortest paths to every other qubit on the chip. Furthermore, the staged extension strategy employed for the Rigetti architecture was unnecessary for this phase. This full-depth execution was feasible due to a combination of two factors: a significantly lower total number of possible shortest paths (a direct consequence of the IBM heavy-hex topology) and a different cloud pricing model that allowed for the unconstrained execution of all routes.
{Figure~\ref{fig:fez_swap_dist} shows return-score distributions for four initial qubits; the remaining four appear in Appendix~\ref{sec:fez_supp_chart}. For qubit 140, the mean return score is below $2/3$ at every evaluated swap distance from 23 to 32, and also at distance 21 before rising above it at 22. At distance 21, fourteen of the 25 evaluated route circuits have return score above $2/3$.} The rule remains strictly $\widehat{\mathcal{F}}>2/3$; the binomial standard error near the cutoff is approximately $0.021$, so nearby classifications are sensitive to finite-shot variation.

\begin{figure*}[!t]
    \centering
    \includegraphics[page=7,trim=79bp 0bp 78.75bp 0bp,clip,width=\linewidth]{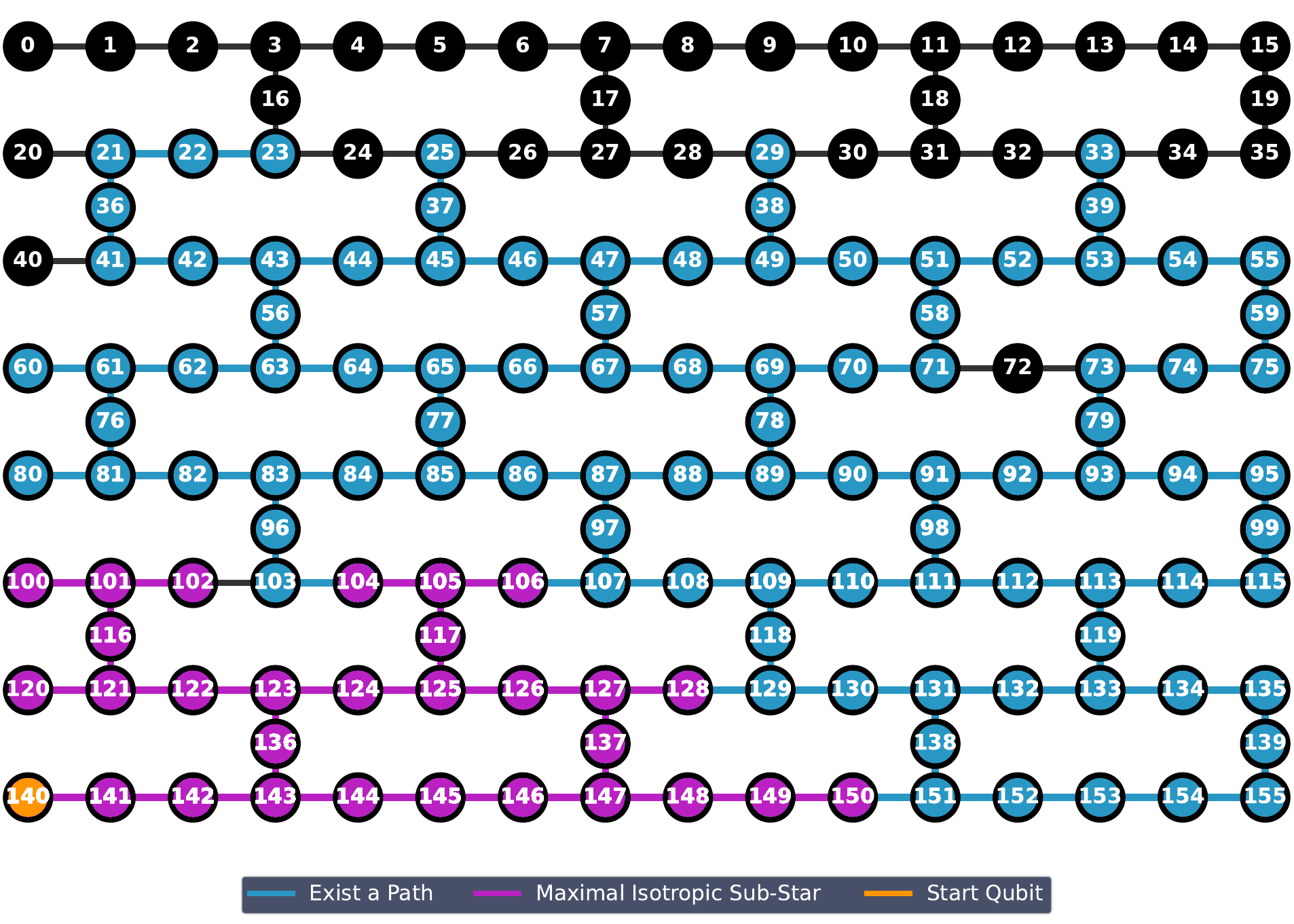}
    \caption{Illustration of the results for \textit{initial qubit} 140 (orange) on the chip connectivity map, light blue color marks all qubits and couplers for which a successful path exists from the initial qubit to a target qubit that contains this qubit or coupler. Magenta marks the \textit{Maximal Isotropic Sub-star} as defined in section~\ref{sec:experimental_design}}
    
    \label{fig:fez_q140_spider_chart}    
\end{figure*}

Figure~\ref{fig:fez_q140_spider_chart} illustrates these results overlaid on the QPU's connectivity map. The magenta nodes in this graph represent one of the empirical answers to our research question: the \textit{Maximal Isotropic Sub-star}. To understand the boundaries of this specific case, we direct the reader's attention to the coupler between qubit 102 and 103, which is colored black.
No evaluated successful path includes that coupler. This identifies a boundary in the observed route map, without isolating an error in the coupler from the rest of the circuit. The measured isotropic radius is ten: all evaluated shortest routes from qubit 140 within that radius pass the empirical prefix rule. A route from qubit 140 to qubit 23 also passes that rule at swap distance 27. These quantities summarize the evaluated circuits under the execution and calibration conditions of the experiment.

\begin{figure*}
    \centering
    \includegraphics[page=8,width=\linewidth]{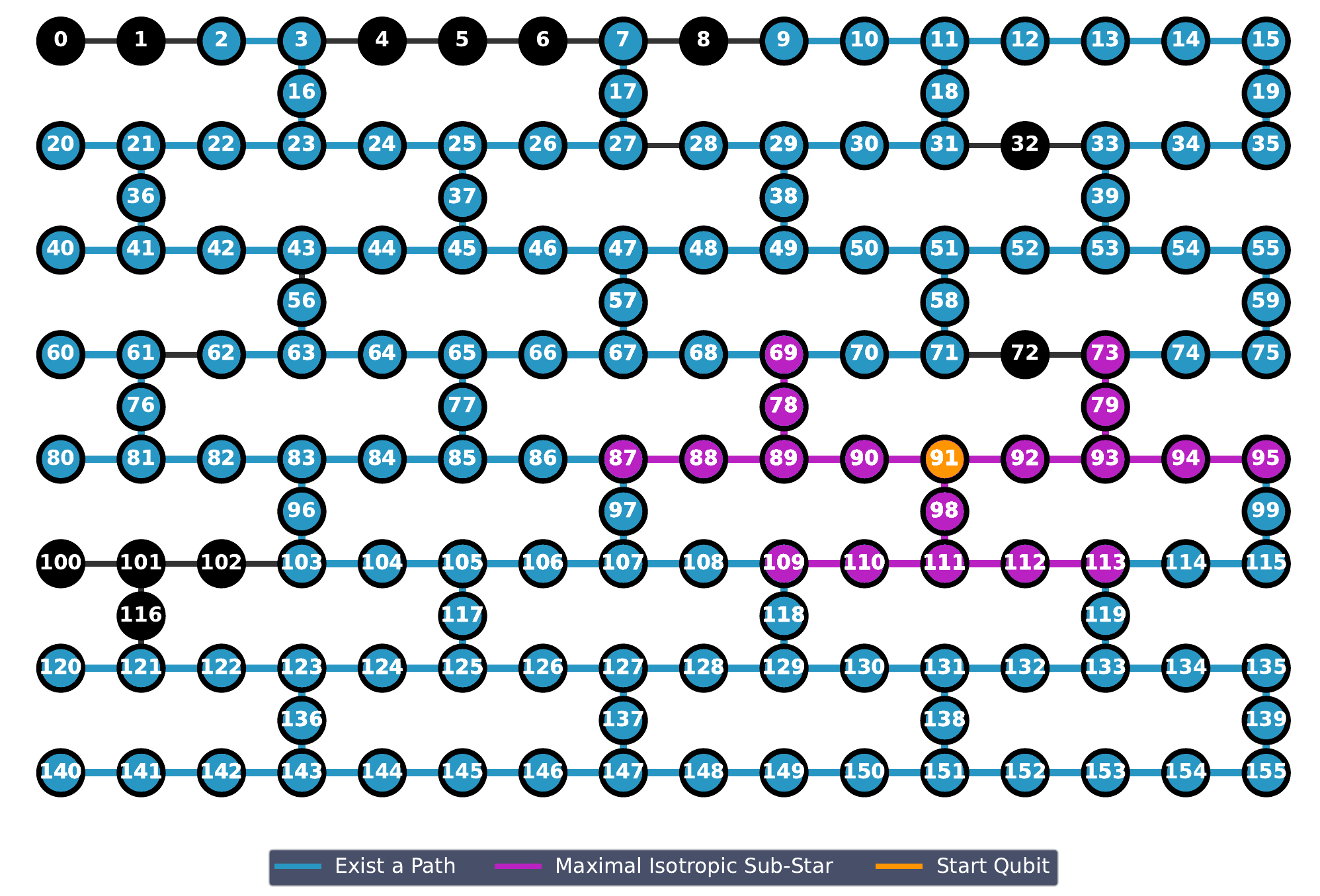}
    \caption{Results of the 'do-nothing' protocol originating from \textit{initial qubit} 91 (orange) on the IBM Fez QPU. This centrally located qubit is presented to provide a direct topological comparison to the central qubits evaluated on the Rigetti processor. The magenta nodes represent the Maximal Isotropic Sub-Star, while light blue nodes signify all targets accessible via at least one successful path.}
    \label{fig:fez_q91_spider_chart}
\end{figure*}

Figure~\ref{fig:fez_q91_spider_chart} shows the corresponding spatial map for the centrally located initial qubit 91, complementing the high-performing qubit-140 case.

The Fez data thus show a finite region of uniform above-cutoff performance together with a longer identified successful route; they do not by themselves attribute the observed boundary to the heavy-hex topology.

\section{Comparison}
We compare the evaluated IBM and Rigetti route families using the same empirical return-score cutoff in the \textit{do-nothing} benchmark.
The route populations are not sampled symmetrically: the IBM statistics include all evaluated shortest paths under the stated routing rule, whereas the later-distance Rigetti statistics are conditioned on survival through the staged filter. For this reason, mean return score at a fixed distance is not used as a like-for-like cross-platform ranking. The primary comparison is restricted to the two explicitly defined route-set-conditional spatial metrics.
Our evaluation reveals distinct architectural paradigms and performance trade-offs.
The Rigetti Cepheus-1-108Q hardware, which offers much more routing flexibility than the evaluated IBM QPUs, demonstrated highly path-dependent performance, identifying successful routes at swap distances of eight and six from selected initial qubits.
While the selected IBM results demonstrate strong isotropic performance and a long successful route for qubit 140, the heavy-hex lattice topology inherently offers fewer alternative routing choices.

\begin{table*}[htbp]
    \centering
    \begin{tabular}{@{}l c c c@{}}
        \toprule
        \textbf{QPU} & \textbf{\makecell{Max Isotropic\\Radius}} & \textbf{\makecell{Max Successful\\Distance}} & \textbf{\makecell{Best Long\\Path}} \\
        \midrule
        IBM (Fez) & 
        10 & 
        27 & 
        $140 \rightarrow 23$ \\
        
        Rigetti Cepheus-1-108Q & 
        1 & 
        8 & 
        $107 \rightarrow 75 \text{ and } 67$ \\
        \bottomrule
    \end{tabular}
    
    \caption{Summary of spatial benchmarking metrics across the presented QPUs, aggregated over routes originating from qubit 140 for IBM and 107 for Rigetti. The table details core spatial metrics extracted for each architecture: the maximal isotropic boundaries and the longest successful paths identified in the evaluated route set.}
    \label{tab:qpu_summary}
\end{table*}

The Cepheus connectivity provides more alternative topological routes. The observed separation between mean and maximum return scores shows that its evaluated routes have heterogeneous performance.

Within the two highlighted cases, Fez qubit 140 attains a larger empirical isotropic radius and a longer identified successful route than the best displayed Rigetti initial qubit. Cepheus, however, supplies a much larger set of alternative topological routes, of which only a staged subset was evaluated at larger distances. The data therefore support a contrast between uniform reliability and routing flexibility, but they do not support a platform-wide superiority claim or an extrapolation to untested Rigetti routes.

\section{Conclusions}\label{sec:conclusions}
Central to our methodology is a common empirical rule: a route and all its evaluated prefixes must have return score above $2/3$. This cutoff classifies the measured circuits; it does not certify above-classical unknown-state transfer for the circuit ordering used here. Finite-shot estimates near the cutoff remain uncertainty-sensitive.
By shifting the benchmarking focus from abstract error rates to functional, protocol-based success, this binary approach provides a more interpretable characterization of complex and distinct hardware grids.
Although our present investigation is focused solely on the \textit{do-nothing} protocol, 
this deliberate choice highlights a crucial hardware reality: current QPUs still struggle to execute even seemingly trivial operations reliably. This work demonstrates that achieving success in \textit{do-nothing}, the almost simplest protocol among the protocol suite proposed in~\cite{meirom2025, mayo2026_1}, is highly non-trivial. We believe that this fact underscores the motivation for spatial benchmarking and the rigorous \textit{Benchmarking via Protocols} of fundamental hardware capabilities. Furthermore, it highlights the necessity of our proposed metrics: the \textit{Maximal Isotropic Sub-star} and the \textit{Maximal Successful Length} metric. Standard gate fidelities fail to capture topological routing limitations; these two metrics are essential to quantify the difference between a processor's uniform, isotropic reliability and the existence of selected long routes that pass the empirical rule.

\begingroup
\widowpenalty=10000
The broader \textit{Benchmarking via Protocols} framework includes teleportation, superdense coding, and transmission of entangled pairs. A classical-threshold certification claim for any such protocol requires a bound valid for its actual circuit, input ensemble, and measured observable, in addition to control of finite-shot uncertainty. The present return-score analysis makes no such certification claim.
The power and scalability of this \textit{benchmarking via protocols} approach are further demonstrated across various hardware paradigms in related work, which utilize it for a deep comparative analysis of IBM superconducting QPUs~\cite{mayo2026_1} and also for Ion-Trap QPU technology~\cite{mayo2026_2}.
By treating functional protocol survival as a complementary metric, these foundational frameworks establish a uniform, hardware-agnostic baseline essential for evaluating the next generation of scalable quantum processors.
\par
\endgroup

While this study focuses exclusively on the \textit{do-nothing} state-transfer task, a natural direction for future research is to extend this spatial benchmarking framework to characterize more general multi-qubit protocols~\cite{meirom2025, mayo2026_2}.

\mainthree{Within its current scope, this study has five important limitations. First, the return score is not a certification of unknown-state transfer, and the available record does not specify the preparation ensemble or its averaging. Second, it evaluates only the \textit{do-nothing} state-transfer task and therefore does not characterize more general multi-qubit protocols. Third, the IBM and Rigetti route populations were generated by different search procedures, and the highlighted initial qubits are not a symmetric random sample. Fourth, the binary labels are computed from 512-shot empirical estimates and are sensitive near the $2/3$ boundary. Fifth, every hardware result is a temporal snapshot exposed to calibration drift. The reported isotropic radii and maximal identified paths should therefore be read as summaries of the recorded route families and execution windows, not as static properties of an entire processor family.}

The central conclusion is that a single average can conceal two operationally different properties: the size of a region in which every evaluated shortest route succeeds, and the existence of selected long routes that succeed despite poor surrounding performance. Reporting both views yields a more informative route-resolved diagnostic for noise-aware placement and routing than either a chip-wide mean or a best-path result alone. This conclusion is limited to the tested protocol, route sets, and execution snapshots.

The retained experimental records include route lists, initial-qubit labels, raw measurement counts, and derived return scores. This article does not provide a public data or code repository or a complete circuit-and-analysis archive. The retained records do not specify the preparation-unitary distribution or the number of independent preparations per route.

\section{Acknowledgment}
We thank both IBM Quantum and Rigetti Computing for their outstanding commitment to hardware transparency and direct qubit control options. The ability to exert direct, low-level control over qubits and couplers across their cloud platforms was essential to the design, execution, and success of this research project.

\printbibliography

@article{Arute_2019,
   title={Quantum supremacy using a programmable superconducting processor},
   volume={574},
   ISSN={1476-4687},
   url={http://dx.doi.org/10.1038/s41586-019-1666-5},
   DOI={10.1038/s41586-019-1666-5},
   number={7779},
   journal={Nature},
   publisher={Springer Science and Business Media LLC},
   author={Arute, Frank and Arya, Kunal and Babbush, Ryan and Bacon, Dave and Bardin, Joseph C. and others},
   year={2019},
   month=oct, pages={505–510} }

@misc{mayo2026_2,
      title={Benchmarking Quantum Computers via Protocols -- Comparing Superconducting and Ion-Trap Quantum Technology}, 
      author={Nitay Mayo and Tal Mor and Yossi Weinstein},
      year={2026},
      eprint={2603.27397},
      archivePrefix={arXiv},
      primaryClass={quant-ph},
      url={https://arxiv.org/abs/2603.27397}, 
}

@misc{meirom2025,
      title={Benchmarking quantum computers via protocols}, 
      author={Dekel Meirom and Tal Mor and Yossi Weinstein},
      year={2025},
      eprint={2505.12441},
      archivePrefix={arXiv},
      primaryClass={quant-ph},
      url={https://arxiv.org/abs/2505.12441}, 
}

@misc{mayo2026_1,
      title={Benchmarking Quantum Computers via Protocols, Comparing IBM's Heron vs IBM's Eagle}, 
      author={Nitay Mayo and Tal Mor and Yossi Weinstein},
      year={2026},
      eprint={2603.04377},
      archivePrefix={arXiv},
      primaryClass={quant-ph},
      url={https://arxiv.org/abs/2603.04377}, 
}

@article{Massar_Popescu_1995,
  title = {Optimal Extraction of Information from Finite Quantum Ensembles},
  author = {Massar, S. and Popescu, S.},
  journal = {Phys. Rev. Lett.},
  volume = {74},
  issue = {8},
  pages = {1259--1263},
  numpages = {0},
  year = {1995},
  month = {2},
  publisher = {American Physical Society},
  doi = {10.1103/PhysRevLett.74.1259},
  url = {https://link.aps.org/doi/10.1103/PhysRevLett.74.1259}
}

@article{kim2023evidence,
  title={Evidence for the utility of quantum computing before fault tolerance},
  author={Kim, Y. and Eddins, A. and Anand, S. and others},
  journal={Nature},
  volume={618},
  pages={500-505},
  year={2023},
  doi={10.1038/s41586-023-06096-3},
  publisher={Nature Publishing Group}
}

@manual{rigetti_cepheus_manual,
    title        = {Cepheus-1-108Q Spec Sheet},
    organization = {Rigetti Computing},
    year         = {2026}, 
    url          = {https://qcs.rigetti.com/static/pdf/108Q%20Spec%20Sheet%20-%20April%202026.pdf},
    note         = {Accessed: 2026-06-14}
}

@misc{norris2025performancecharacterizationmultimodulequantum,
      title={Performance Characterization of a Multi-Module Quantum Processor with Static Inter-Chip Couplers}, 
      author={Graham J. Norris and Kieran Dalton and Dante Colao Zanuz and Alexander Rommens and Alexander Flasby and Mohsen Bahrami Panah and François Swiadek and Colin Scarato and Christoph Hellings and Jean-Claude Besse and Andreas Wallraff},
      year={2025},
      eprint={2503.12603},
      archivePrefix={arXiv},
      primaryClass={quant-ph},
      url={https://arxiv.org/abs/2503.12603}, 
}

@misc{zhao2026longrangetunablecouplermodular,
      title={Long-range tunable coupler for modular fluxonium quantum processors}, 
      author={Peng Zhao and Peng Xu and Zheng-Yuan Xue},
      year={2026},
      eprint={2604.12261},
      archivePrefix={arXiv},
      primaryClass={quant-ph},
      url={https://arxiv.org/abs/2604.12261}, 
}

@misc{gold2021entanglementseparatesilicondies,
      title={Entanglement Across Separate Silicon Dies in a Modular Superconducting Qubit Device}, 
      author={Alysson Gold and JP Paquette and Anna Stockklauser and Matthew J. Reagor and M. Sohaib Alam and et al. and Chad Rigetti},
      year={2021},
      eprint={2102.13293},
      archivePrefix={arXiv},
      primaryClass={quant-ph},
      url={https://arxiv.org/abs/2102.13293}, 
}

@misc{murali2019noiseadaptivecompilermappingsnoisy,
      title={Noise-Adaptive Compiler Mappings for Noisy Intermediate-Scale Quantum Computers}, 
      author={Prakash Murali and Jonathan M. Baker and Ali Javadi Abhari and Frederic T. Chong and Margaret Martonosi},
      year={2019},
      eprint={1901.11054},
      archivePrefix={arXiv},
      primaryClass={quant-ph},
      url={https://arxiv.org/abs/1901.11054}, 
}

@article{Sarovar_2020,
   title={Detecting crosstalk errors in quantum information processors},
   volume={4},
   ISSN={2521-327X},
   url={http://dx.doi.org/10.22331/q-2020-09-11-321},
   DOI={10.22331/q-2020-09-11-321},
   journal={Quantum},
   publisher={Verein zur Forderung des Open Access Publizierens in den Quantenwissenschaften},
   author={Sarovar, Mohan and Proctor, Timothy and Rudinger, Kenneth and Young, Kevin and Nielsen, Erik and Blume-Kohout, Robin},
   year={2020},
   month=sep, pages={321} }

@inproceedings{Ding_2020,
   title={Systematic Crosstalk Mitigation for Superconducting Qubits via Frequency-Aware Compilation},
   url={http://dx.doi.org/10.1109/MICRO50266.2020.00028},
   DOI={10.1109/micro50266.2020.00028},
   booktitle={2020 53rd Annual IEEE/ACM International Symposium on Microarchitecture (MICRO)},
   publisher={IEEE},
   author={Ding, Yongshan and Gokhale, Pranav and Lin, Sophia Fuhui and Rines, Richard and Propson, Thomas and Chong, Frederic T.},
   year={2020},
   month=Oct, pages={201–214} }

@article{Proctor_2021,
   title={Measuring the capabilities of quantum computers},
   volume={18},
   ISSN={1745-2481},
   url={http://dx.doi.org/10.1038/s41567-021-01409-7},
   DOI={10.1038/s41567-021-01409-7},
   number={1},
   journal={Nature Physics},
   publisher={Springer Science and Business Media LLC},
   author={Proctor, Timothy and Rudinger, Kenneth and Young, Kevin and Nielsen, Erik and Blume-Kohout, Robin},
   year={2021},
   month=Dec, pages={75–79} }

@article{Preskill_2018,
   title={Quantum Computing in the NISQ era and beyond},
   volume={2},
   ISSN={2521-327X},
   url={http://dx.doi.org/10.22331/q-2018-08-06-79},
   DOI={10.22331/q-2018-08-06-79},
   journal={Quantum},
   publisher={Verein zur Forderung des Open Access Publizierens in den Quantenwissenschaften},
   author={Preskill, John},
   year={2018},
   month=Aug, pages={79} }

@article{Pfaff_2014,
   title={Unconditional quantum teleportation between distant solid-state quantum bits},
   volume={345},
   ISSN={1095-9203},
   url={http://dx.doi.org/10.1126/science.1253512},
   DOI={10.1126/science.1253512},
   number={6196},
   journal={Science},
   publisher={American Association for the Advancement of Science (AAAS)},
   author={Pfaff, W. and Hensen, B. J. and Bernien, H. and van Dam, S. B. and Blok, M. S. and Taminiau, T. H. and Tiggelman, M. J. and Schouten, R. N. and Markham, M. and Twitchen, D. J. and Hanson, R.},
   year={2014},
   month=Aug, pages={532–535} }

@article{marquez_teleportation_benchmark_2025,
  author       = {M{\'a}rquez, Cristian and Sierra-Sosa, Daniel and Garc{\'e}s, Kelly},
  title        = {A Teleportation Protocol Variant for Single-{QPU} Benchmarking},
  journal      = {IEEE Access},
  volume       = {13},
  pages        = {209266--209281},
  year         = {2025},
  doi          = {10.1109/ACCESS.2025.3639914}
}

@article{pelofske_qv_2022,
  author       = {Pelofske, Elijah and B{\"a}rtschi, Andreas and Eidenbenz, Stephan},
  title        = {Quantum Volume in Practice: What Users Can Expect From {NISQ} Devices},
  journal      = {IEEE Transactions on Quantum Engineering},
  volume       = {3},
  year         = {2022},
  doi          = {10.1109/TQE.2022.3184764}
}

@article{lozano_layer_fidelity_2026,
  author       = {Lozano Palacio, Maria Jose and Nayfeh, Hasan and Ware, Matthew and McKay, David C.},
  title        = {Parameter Analysis and Optimization of Layer Fidelity for Quantum Processor Benchmarking at Scale},
  journal      = {IEEE Transactions on Quantum Engineering},
  year         = {2026},
  doi          = {10.1109/TQE.2026.3668098}
}

@inproceedings{suau_single_qubit_2023,
  author    = {Suau, Adrien and Nelson, Jon and Vuffray, Marc and Lokhov, Andrey Y. and Cincio, Lukasz and Coffrin, Carleton},
  title     = {Single-Qubit Cross Platform Comparison of Quantum Computing Hardware},
  booktitle = {2023 IEEE International Conference on Quantum Computing and Engineering (QCE)},
  year      = {2023},
  doi       = {10.1109/QCE57702.2023.00155}
}

@article{lubinski_app_bench_2023,
  author       = {Lubinski, Thomas and Johri, Sonika and Varosy, Paul and Coleman, Jeremiah and Zhao, Luning and Necaise, Jason and Baldwin, Charles H. and Mayer, Karl and Proctor, Timothy},
  title        = {Application-Oriented Performance Benchmarks for Quantum Computing},
  journal      = {IEEE Transactions on Quantum Engineering},
  volume       = {4},
  year         = {2023},
  doi          = {10.1109/TQE.2023.3253761}
}

@inproceedings{tomesh_supermarq_2022,
  author    = {Tomesh, Teague and Gokhale, Pranav and Omole, Victory and Ravi, Gokul Subramanian and Smith, Kaitlin N. and Viszlai, Joshua and Wu, Xin-Chuan and Hardavellas, Nikos and Martonosi, Margaret R. and Chong, Frederic T.},
  title     = {{SupermarQ}: A Scalable Quantum Benchmark Suite},
  booktitle = {2022 IEEE International Symposium on High-Performance Computer Architecture (HPCA)},
  year      = {2022},
  doi       = {10.1109/HPCA53966.2022.00050}
}

@article{hines_fullstack_2024,
  author       = {Hines, Jordan and Proctor, Timothy},
  title        = {Scalable Full-Stack Benchmarks for Quantum Computers},
  journal      = {IEEE Transactions on Quantum Engineering},
  year         = {2024},
  doi          = {10.1109/TQE.2024.3404502}
}

@article{li_qknob_2025,
  author       = {Li, Sanjiang and Zhou, Xiangzhen and Feng, Yuan},
  title        = {Benchmarking Quantum Circuit Transformation With {QKNOB} Circuits},
  journal      = {IEEE Transactions on Quantum Engineering},
  volume       = {6},
  year         = {2025},
  doi          = {10.1109/TQE.2025.3527399}
}

@article{zhukov_protocol_benchmark_2019,
  author       = {Zhukov, A. A. and Kiktenko, E. O. and Elistratov, A. A. and Pogosov, W. V. and Lozovik, Yu. E.},
  title        = {Quantum Communication Protocols as a Benchmark for Programmable Quantum Computers},
  journal      = {Quantum Information Processing},
  volume       = {18},
  number       = {1},
  pages        = {31},
  year         = {2019},
  doi          = {10.1007/s11128-018-2144-y}
}
\clearpage
\appendix
\setcounter{figure}{0}
\renewcommand{\thefigure}{S\arabic{figure}}
\section{Additional Return Score as a Function of Swap Distance Chart for Fez}\label{sec:fez_supp_chart}
As illustrated in Fig.~\ref{fig:fez_fid_vs_swap_dist_chart_2}, qubit 0 (represented by the blue scatter points) initially appears to be a strong starting node, dominating the extended swap distances beyond 23. However, our prefix-closed success criteria restrict its overall routing viability. While some longer-route circuits have relatively high return scores, a high portion of the shorter, intermediate paths originating from qubit 0---particularly between swap distances of 9 and 23---fell below the empirical return-score cutoff of $\mathcal{F}=2/3$. Because the prefix-closed constraint mandates that every intermediate sub-path must independently satisfy the success criteria to keep a route open, below-cutoff prefixes exclude affected routes from the successful set, disqualifying affected long-distance routes from initial qubit 0 despite its apparent long-range performance.

\begin{figure*}[!t]
  \centering
  \includegraphics[page=9,width=\linewidth]{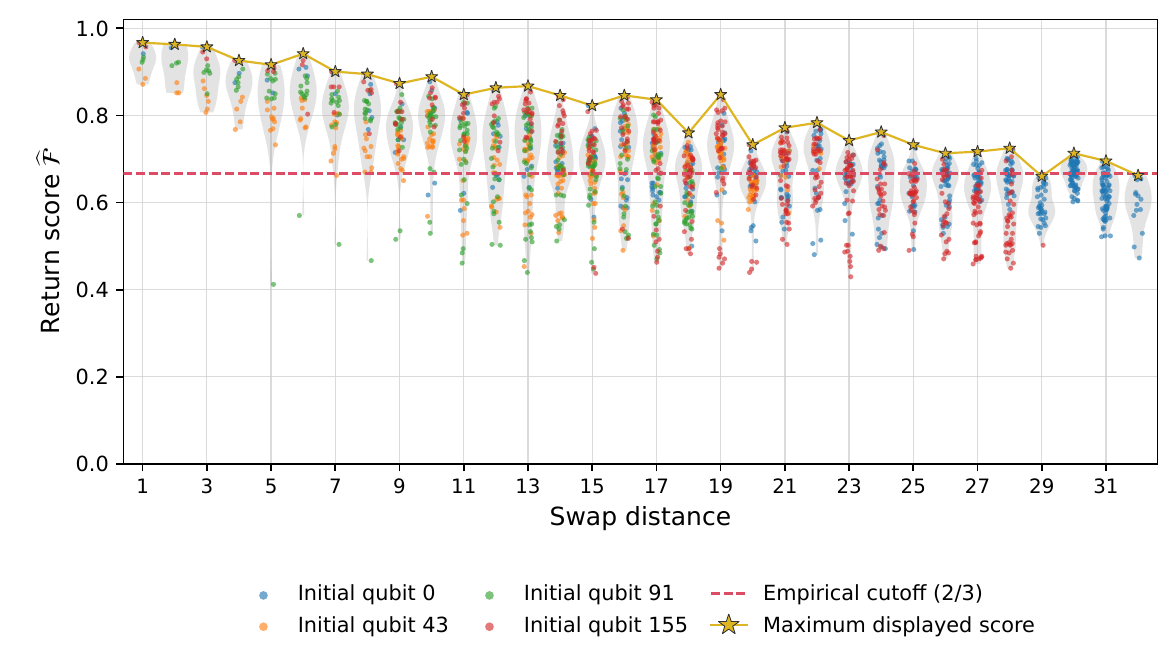}
  \caption{Return score as a function of swap distance on the IBM Fez processor. Grey violins summarize the recorded route scores for the four initial qubits listed in this panel, while scatter points detail individual route outcomes color-coded by a representative subset of four initial qubits (0, 43, 91 and 155). The red dashed line marks the empirical cutoff ($\mathcal{F}=2/3$), and the yellow trajectory tracks the maximum return score at each depth.}
  \label{fig:fez_fid_vs_swap_dist_chart_2}
\end{figure*}

\section{Supplementary data for IBM's QPU}\label{sec:ibm_appendix}
The following spatial connectivity maps show the evaluated initial qubits listed in their captions across the tested IBM quantum processors. The initial qubit is orange, successful-route components are light blue, and the Maximal Isotropic Sub-star is magenta, as in the main manuscript.

\subsection{Fez}
Figures~\ref{fig:fez_q0_supp_map}--\ref{fig:fez_q155_supp_map} show the maps for initial qubits 0, 15, 43, 51, 83 and 155.

\subsection{Kingston}
Figures~\ref{fig:kingston_q0_supp_map}--\ref{fig:kingston_q155_supp_map} show the maps for initial qubits 0, 15, 43, 51, 83, 91, 140 and 155.

\subsection{Marrakesh}
Figures~\ref{fig:marrakesh_q0_supp_map}--\ref{fig:marrakesh_q155_supp_map} show the maps for initial qubits 0, 15, 43, 51, 83, 91, 140 and 155.

\clearpage

\SupplementMap{10}{IBM Fez spatial connectivity map for initial qubit 0.}{fig:fez_q0_supp_map}
\SupplementMap{11}{IBM Fez spatial connectivity map for initial qubit 15.}{fig:fez_q15_supp_map}
\SupplementMap{12}{IBM Fez spatial connectivity map for initial qubit 43.}{fig:fez_q43_supp_map}
\SupplementMap{13}{IBM Fez spatial connectivity map for initial qubit 51.}{fig:fez_q51_supp_map}
\SupplementMap{14}{IBM Fez spatial connectivity map for initial qubit 83.}{fig:fez_q83_supp_map}
\SupplementMap{15}{IBM Fez spatial connectivity map for initial qubit 155.}{fig:fez_q155_supp_map}

\SupplementMap{16}{IBM Kingston spatial connectivity map for initial qubit 0.}{fig:kingston_q0_supp_map}
\SupplementMap{17}{IBM Kingston spatial connectivity map for initial qubit 15.}{fig:kingston_q15_supp_map}
\SupplementMap{18}{IBM Kingston spatial connectivity map for initial qubit 43.}{fig:kingston_q43_supp_map}
\SupplementMap{19}{IBM Kingston spatial connectivity map for initial qubit 51.}{fig:kingston_q51_supp_map}
\SupplementMap{20}{IBM Kingston spatial connectivity map for initial qubit 83.}{fig:kingston_q83_supp_map}
\SupplementMap{21}{IBM Kingston spatial connectivity map for initial qubit 91.}{fig:kingston_q91_supp_map}
\SupplementMap{22}{IBM Kingston spatial connectivity map for initial qubit 140.}{fig:kingston_q140_charts}
\SupplementMap{23}{IBM Kingston spatial connectivity map for initial qubit 155.}{fig:kingston_q155_supp_map}

\SupplementMap{24}{IBM Marrakesh spatial connectivity map for initial qubit 0.}{fig:marrakesh_q0_supp_map}
\SupplementMap{25}{IBM Marrakesh spatial connectivity map for initial qubit 15.}{fig:marrakesh_q15_supp_map}
\SupplementMap{26}{IBM Marrakesh spatial connectivity map for initial qubit 43.}{fig:marrakesh_q43_supp_map}
\SupplementMap{27}{IBM Marrakesh spatial connectivity map for initial qubit 51.}{fig:marrakesh_q51_supp_map}
\SupplementMap{28}{IBM Marrakesh spatial connectivity map for initial qubit 83.}{fig:marrakesh_q83_supp_map}
\SupplementMap{29}{IBM Marrakesh spatial connectivity map for initial qubit 91.}{fig:marrakesh_q91_supp_map}
\SupplementMap{30}{IBM Marrakesh spatial connectivity map for initial qubit 140.}{fig:marrakesh_q140_supp_map}
\SupplementMap{31}{IBM Marrakesh spatial connectivity map for initial qubit 155.}{fig:marrakesh_q155_supp_map}

\begin{figure*}[p]
  \begin{minipage}{\linewidth}
    \section{Additional return-score-versus-distance charts}\label{sec:additional_swap_distance}
    
    Figures~\ref{fig:kingston_swap_distance_1} and~\ref{fig:kingston_swap_distance_2} provide the return-score distributions for Kingston, and Figs.~\ref{fig:marrakesh_swap_distance_1} and~\ref{fig:marrakesh_swap_distance_2} provide the corresponding distributions for Marrakesh. For each processor, the two plots cover the eight evaluated initial qubits in two groups of four, identified in the legends and captions. The return-score and swap-distance axes have the same meanings as in the main manuscript and Fig.~\ref{fig:fez_fid_vs_swap_dist_chart_2}.
  \end{minipage}
  \par\medskip
  \centering
  \includegraphics[page=32,width=\linewidth]{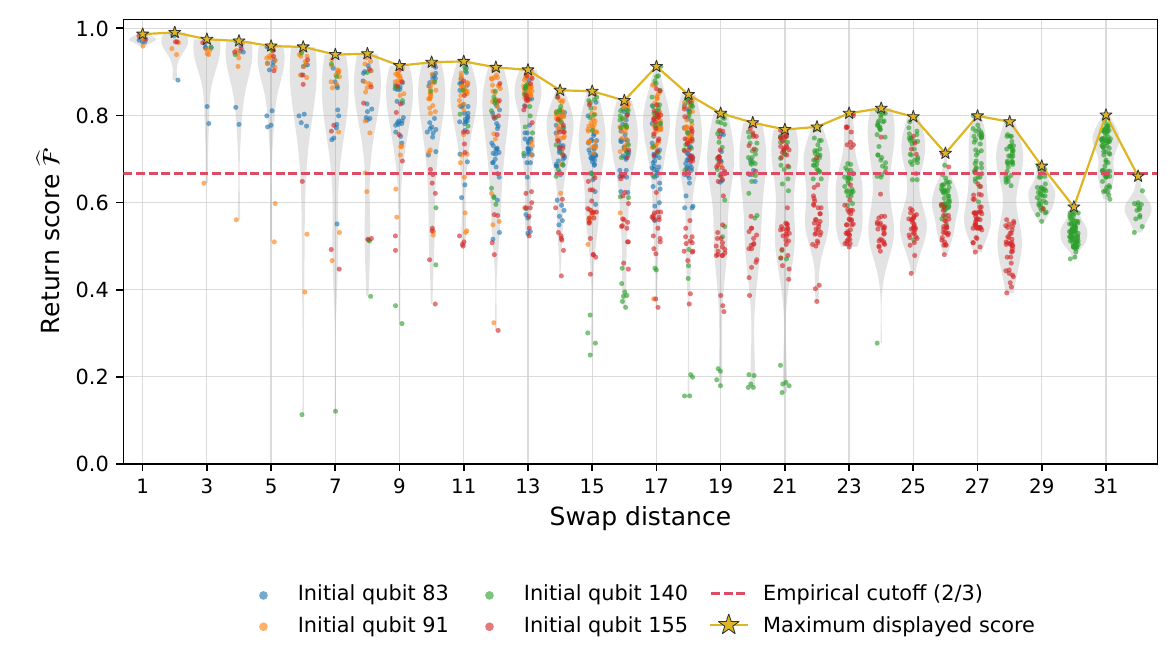}
  \caption{IBM Kingston: return score versus swap distance for initial qubits 83, 91, 140 and 155. Grey violins summarize the recorded route scores for the four initial qubits listed in this panel, and colored points show route outcomes, with the initial qubit identified by the legend. The dashed red line marks $\mathcal{F}=2/3$; the yellow star-marked curve shows the plotted maximum at each swap distance.}
  \label{fig:kingston_swap_distance_1}
\end{figure*}

\begin{figure*}[p]
  \centering
  \includegraphics[page=33,width=\linewidth]{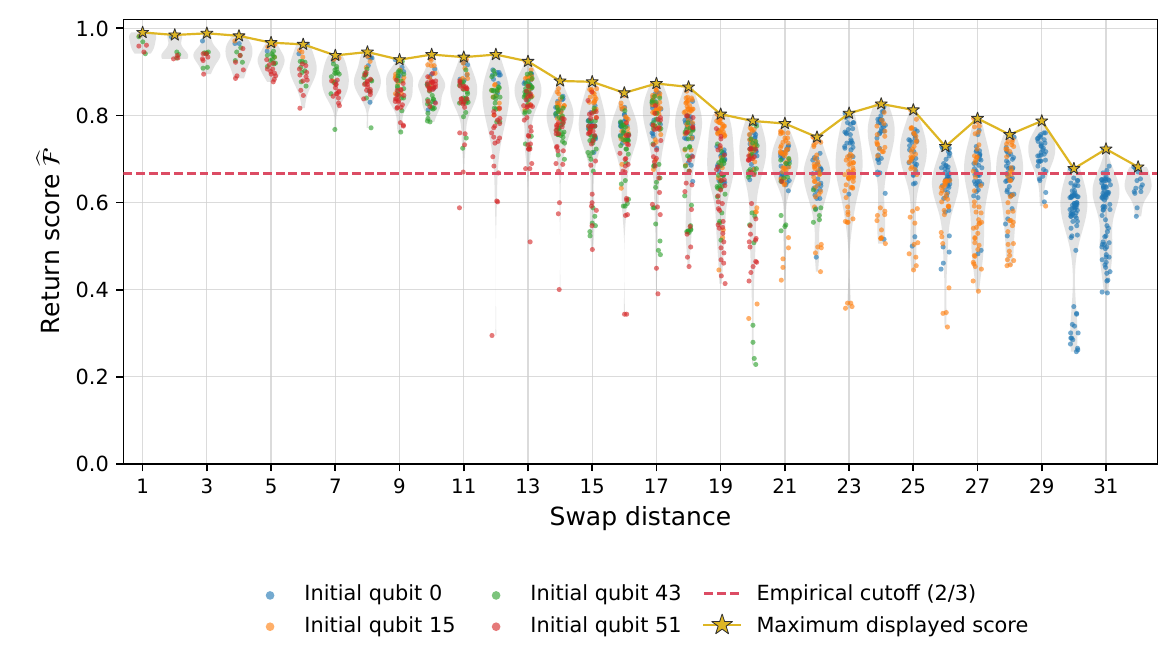}
  \caption{IBM Kingston: return score versus swap distance for initial qubits 0, 15, 43 and 51, with the same plot encodings as in Fig.~\ref{fig:kingston_swap_distance_1}. Together, the two Kingston plots cover all eight evaluated initial qubits.}
  \label{fig:kingston_swap_distance_2}
\end{figure*}
\clearpage

\begin{figure*}[p]
  \centering
  \includegraphics[page=34,width=\linewidth]{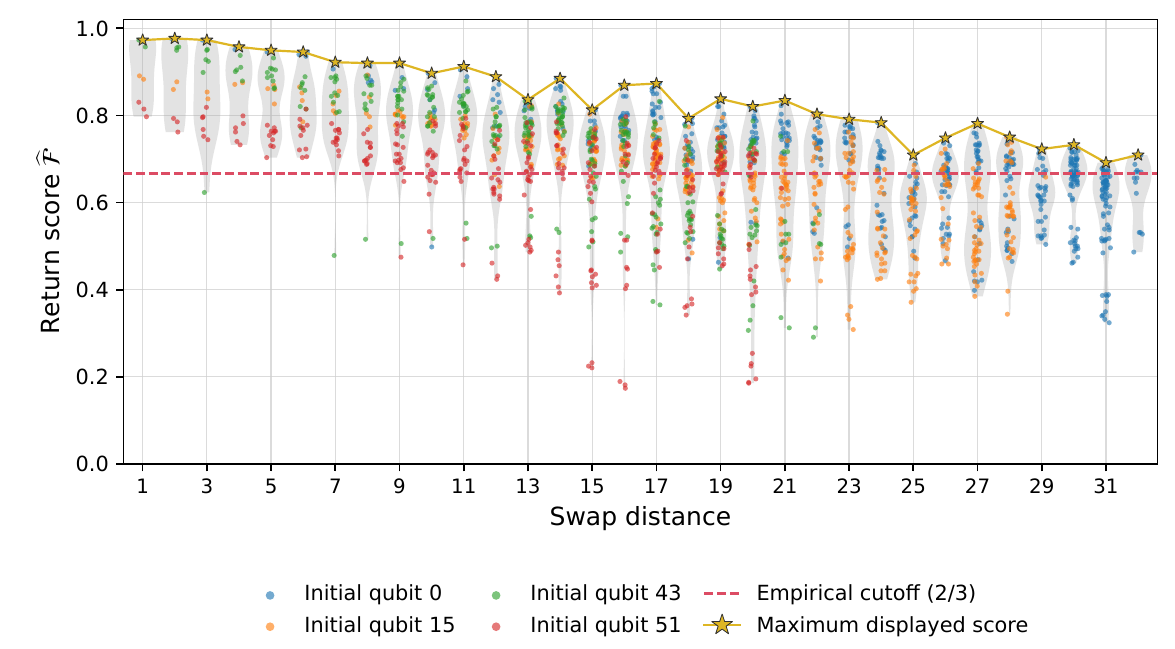}
  \caption{IBM Marrakesh: return score versus swap distance for initial qubits 0, 15, 43 and 51. Grey violins summarize the recorded route scores for the four initial qubits listed in this panel, and colored points show route outcomes, with the initial qubit identified by the legend. The dashed red line marks $\mathcal{F}=2/3$; the yellow star-marked curve shows the plotted maximum at each swap distance.}
  \label{fig:marrakesh_swap_distance_1}
\end{figure*}

\begin{figure*}[p]
  \centering
  \includegraphics[page=35,width=\linewidth]{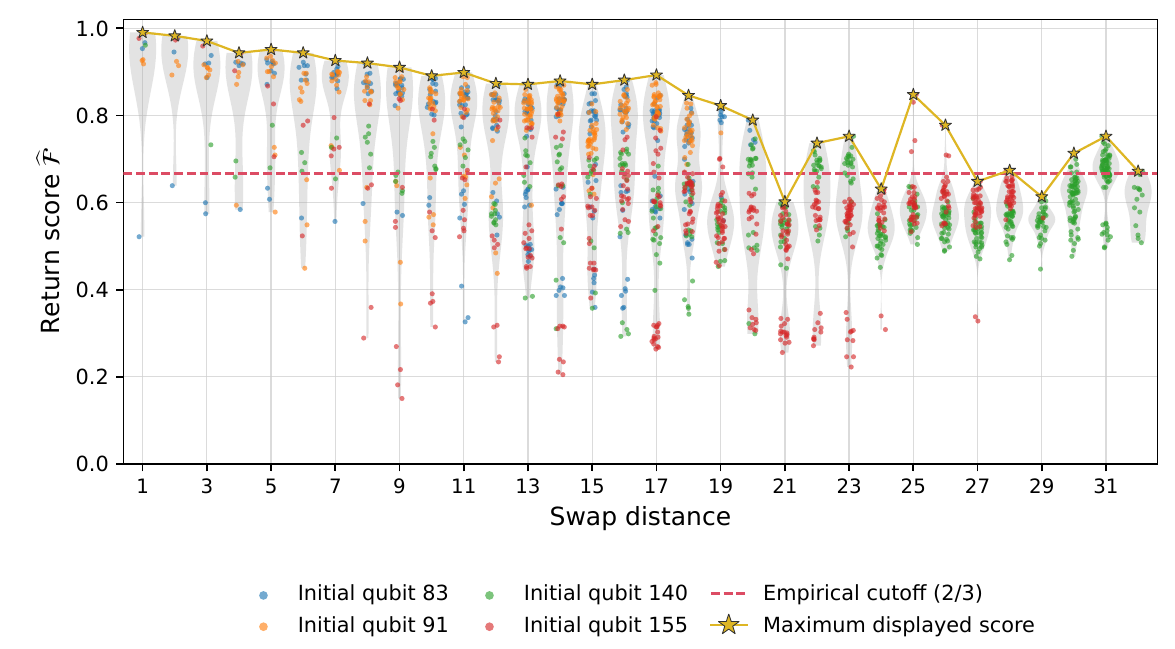}
  \caption{IBM Marrakesh: return score versus swap distance for initial qubits 83, 91, 140 and 155, with the same plot encodings as in Fig.~\ref{fig:marrakesh_swap_distance_1}. Together, the two Marrakesh plots cover all eight evaluated initial qubits.}
  \label{fig:marrakesh_swap_distance_2}
  
\end{figure*}
\clearpage

\clearpage
\section{Detailed plots by initial qubit}\label{sec:detailed_plots}
The following plots provide individual initial-qubit summaries for the evaluated routes, complementing the return-score distributions and spatial maps above.
In addition to the primary work-state metrics discussed in the main text, the return-score charts presented in this section feature supplementary data regarding ancillary qubits (labeled as "Ancillas Min" and "Ancillas Max"). To capture this data, we defined a path-length-dependent mask that selected four specific ancilla qubits situated along the routing trajectory. For paths shorter than four qubits, all participating qubits along the route were measured. These selected ancillas were then measured immediately following the execution of the protocol as supplementary outcome diagnostics.

These ancillary outcomes are supplementary diagnostics. They do not separately identify environmental noise or crosstalk, and we omit them from the main figures to focus on the work-qubit return score.

\begin{figure}[p]
\centering
\begin{subfigure}{0.96\linewidth}
\centering
\includegraphics[width=\linewidth,height=0.34\textheight,keepaspectratio]{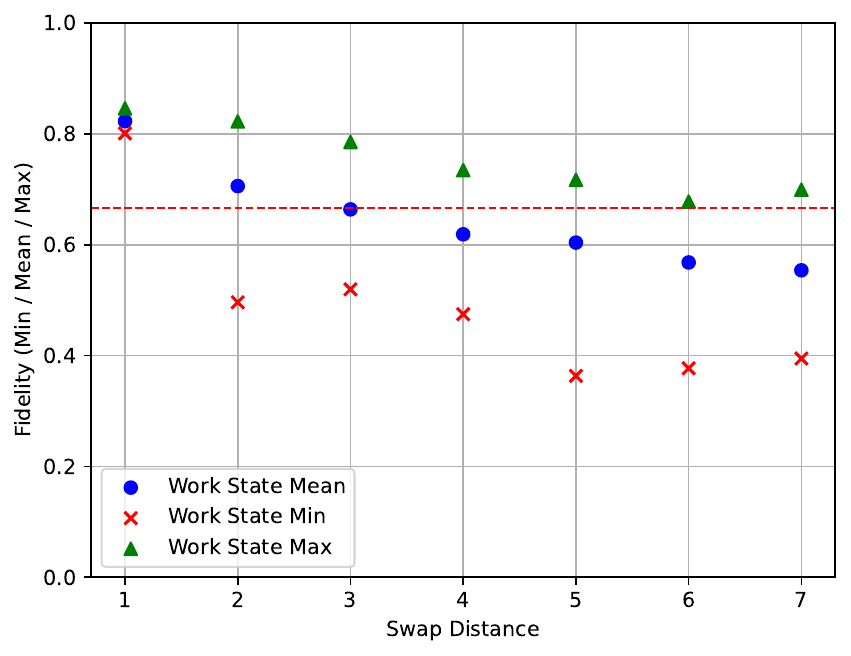}
\caption{Rigetti Cepheus-1-108Q: Mean, Minimum, and Maximum return score as a function of swap distance for paths originating from qubit 67. We note that an extended path of length seven was excluded from the maximal path identification as a result of failing the success criteria defined in Section~\ref{sec:experimental_design}}
\end{subfigure}
\par\medskip
\begin{subfigure}{0.96\linewidth}
\centering
\includegraphics[width=\linewidth,height=0.34\textheight,keepaspectratio]{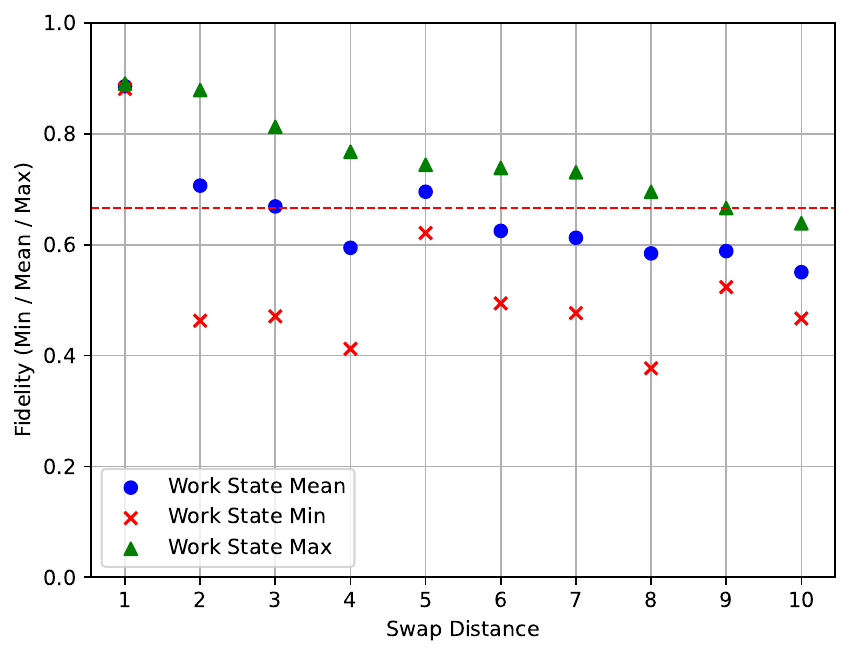}
\caption{Rigetti Cepheus-1-108Q: Mean, Minimum, and Maximum return score as a function of swap distance for paths originating from qubit 107}
\end{subfigure}
\par\medskip
\caption{Detailed return-score-versus-distance summaries for the initial qubits identified in the panels. In these retained plots, work-state \emph{Fidelity} denotes the measured return score; the $2/3$ line is the empirical cutoff.}\label{fig:detailed_1}
\end{figure}
\clearpage
\begin{figure}[p]
\centering
\begin{subfigure}{0.96\linewidth}
\centering
\includegraphics[width=\linewidth,height=0.34\textheight,keepaspectratio]{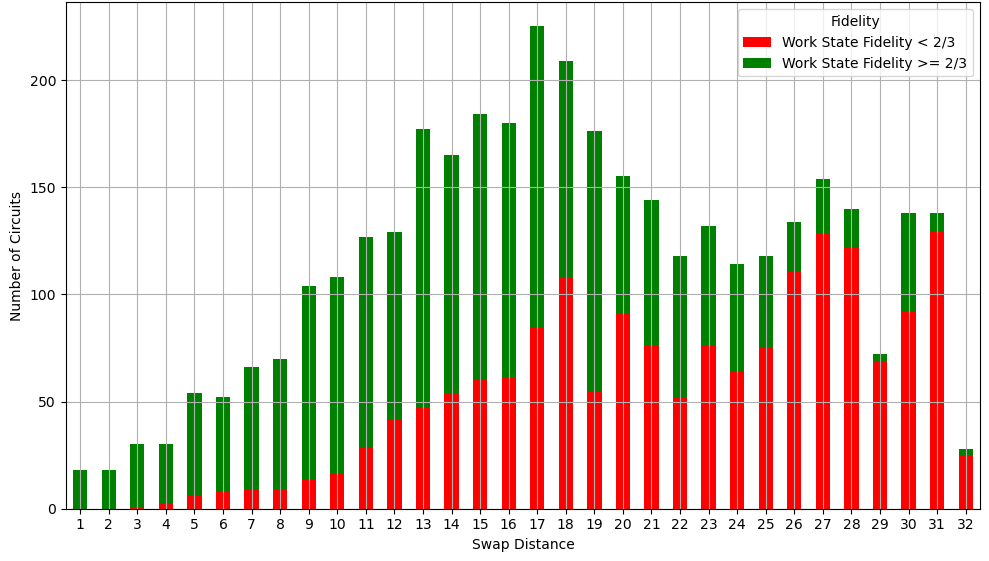}
\caption{IBM Fez: Route circuits with return scores above $2/3$ (green) and below $2/3$ (red), grouped by swap distance. These are circuit-level labels and do not impose the prefix rule. This chart includes all executed Fez route circuits; the next panel shows only qubit 140. With 512 shots, an empirical score cannot equal $2/3$.}
\end{subfigure}
\par\medskip
\begin{subfigure}{0.96\linewidth}
\centering
\includegraphics[width=\linewidth,height=0.34\textheight,keepaspectratio]{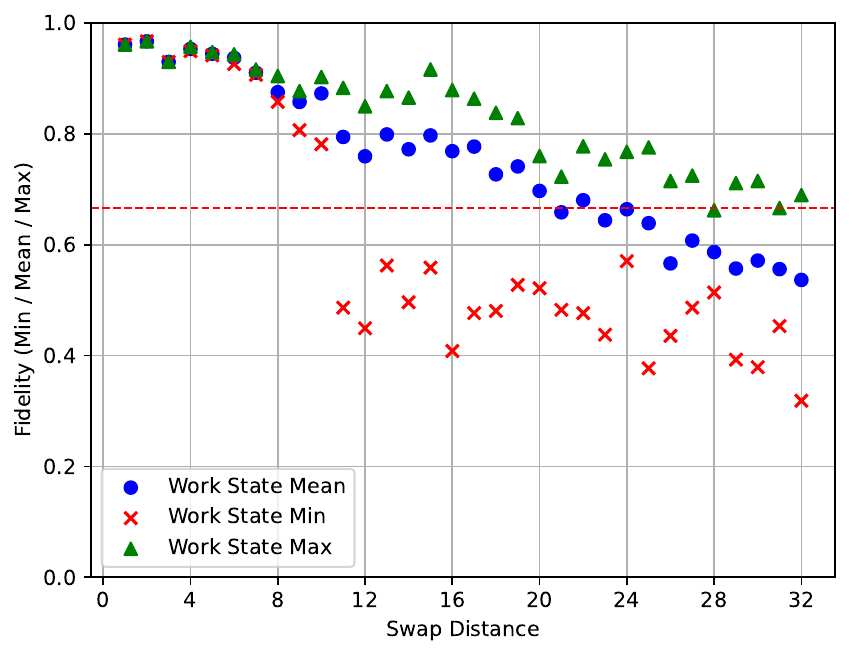}
\caption{IBM Fez: Return score as a function of swap distance for the 'do-nothing' protocol, only for paths originating from qubit 140. The markers (blue, green, and red) denote the mean, maximum, and minimum return scores, respectively. The horizontal dashed line represents the $2/3$ empirical cutoff.}
\end{subfigure}
\par\medskip
\caption{Detailed return-score-versus-distance summaries for the initial qubits identified in the panels. In these retained plots, work-state \emph{Fidelity} denotes the measured return score; the $2/3$ line is the empirical cutoff.}\label{fig:detailed_2}
\end{figure}
\clearpage
\begin{figure}[p]
\centering
\begin{subfigure}{0.96\linewidth}
\centering
\includegraphics[width=\linewidth,height=0.34\textheight,keepaspectratio]{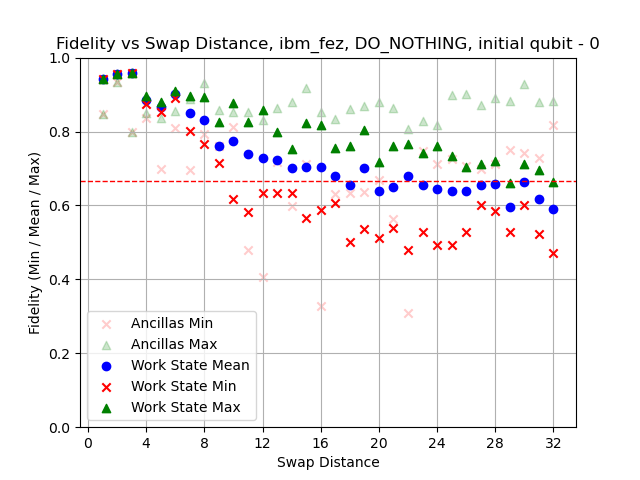}
\caption{IBM Fez: Initial qubit 0.}
\end{subfigure}
\par\medskip
\begin{subfigure}{0.96\linewidth}
\centering
\includegraphics[width=\linewidth,height=0.34\textheight,keepaspectratio]{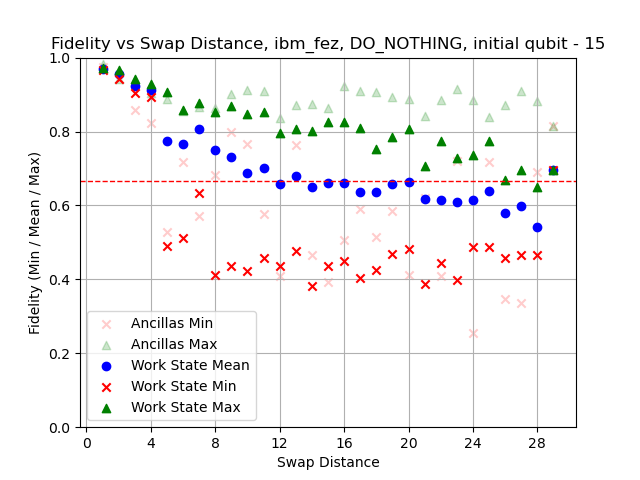}
\caption{IBM Fez: Initial qubit 15.}
\end{subfigure}
\par\medskip
\caption{Detailed return-score-versus-distance summaries for the initial qubits identified in the panels. In these retained plots, work-state \emph{Fidelity} denotes the measured return score; the $2/3$ line is the empirical cutoff.}\label{fig:detailed_3}
\end{figure}
\clearpage
\begin{figure}[p]
\centering
\begin{subfigure}{0.96\linewidth}
\centering
\includegraphics[width=\linewidth,height=0.34\textheight,keepaspectratio]{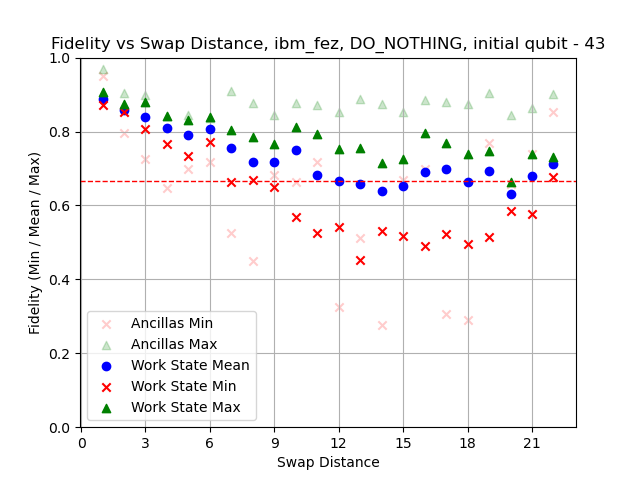}
\caption{IBM Fez: Initial qubit 43.}
\end{subfigure}
\par\medskip
\begin{subfigure}{0.96\linewidth}
\centering
\includegraphics[width=\linewidth,height=0.34\textheight,keepaspectratio]{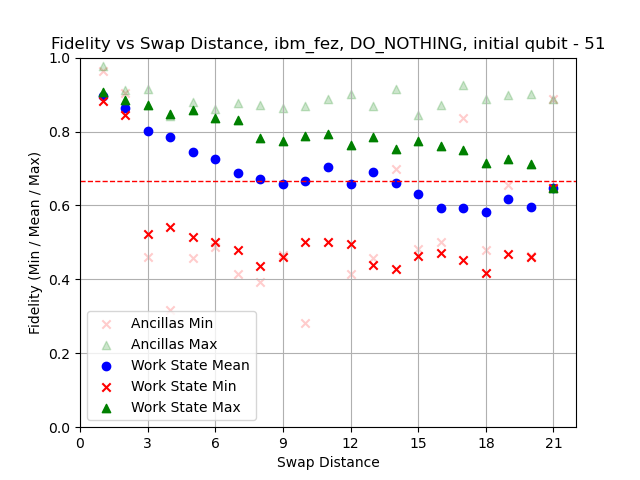}
\caption{IBM Fez: Initial qubit 51.}
\end{subfigure}
\par\medskip
\caption{Detailed return-score-versus-distance summaries for the initial qubits identified in the panels. In these retained plots, work-state \emph{Fidelity} denotes the measured return score; the $2/3$ line is the empirical cutoff.}\label{fig:detailed_4}
\end{figure}
\clearpage
\begin{figure}[p]
\centering
\begin{subfigure}{0.96\linewidth}
\centering
\includegraphics[width=\linewidth,height=0.34\textheight,keepaspectratio]{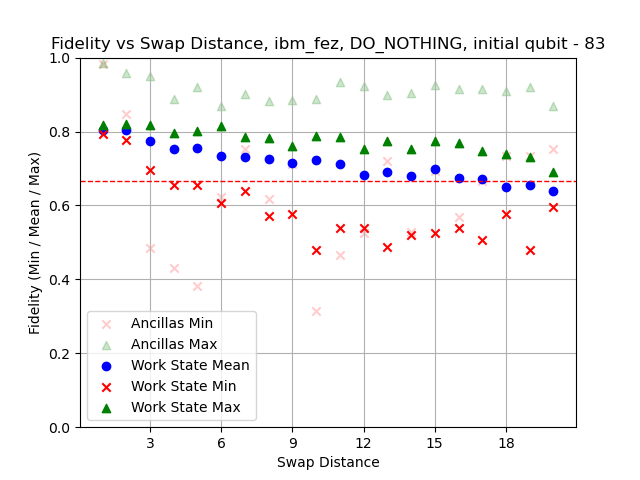}
\caption{IBM Fez: Initial qubit 83.}
\end{subfigure}
\par\medskip
\begin{subfigure}{0.96\linewidth}
\centering
\includegraphics[width=\linewidth,height=0.34\textheight,keepaspectratio]{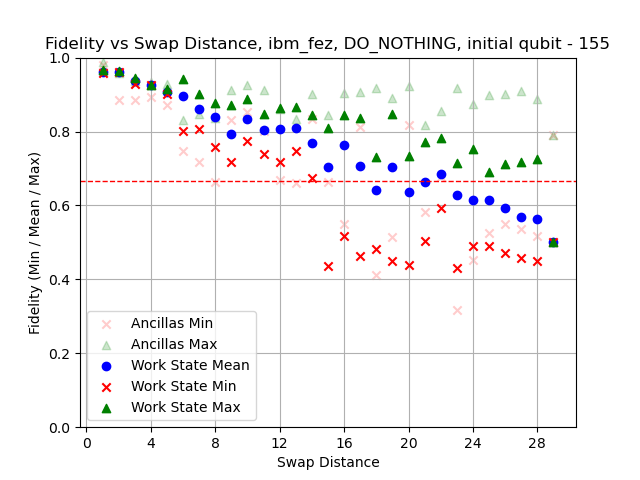}
\caption{IBM Fez: Initial qubit 155.}
\end{subfigure}
\par\medskip
\caption{Detailed return-score-versus-distance summaries for the initial qubits identified in the panels. In these retained plots, work-state \emph{Fidelity} denotes the measured return score; the $2/3$ line is the empirical cutoff.}\label{fig:detailed_5}
\end{figure}
\clearpage
\begin{figure}[p]
\centering
\begin{subfigure}{0.96\linewidth}
\centering
\includegraphics[width=\linewidth,height=0.34\textheight,keepaspectratio]{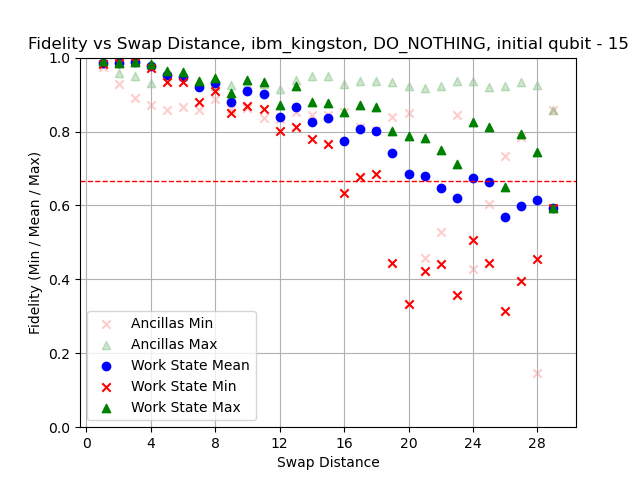}
\caption{IBM Kingston: Initial qubit 15.}
\end{subfigure}
\par\medskip
\begin{subfigure}{0.96\linewidth}
\centering
\includegraphics[width=\linewidth,height=0.34\textheight,keepaspectratio]{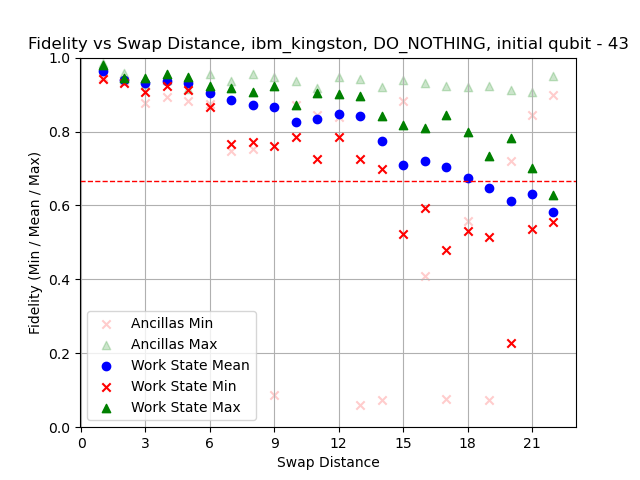}
\caption{IBM Kingston: Initial qubit 43.}
\end{subfigure}
\par\medskip
\caption{Detailed return-score-versus-distance summaries for the initial qubits identified in the panels. In these retained plots, work-state \emph{Fidelity} denotes the measured return score; the $2/3$ line is the empirical cutoff.}\label{fig:detailed_6}
\end{figure}
\clearpage
\begin{figure}[p]
\centering
\begin{subfigure}{0.96\linewidth}
\centering
\includegraphics[width=\linewidth,height=0.34\textheight,keepaspectratio]{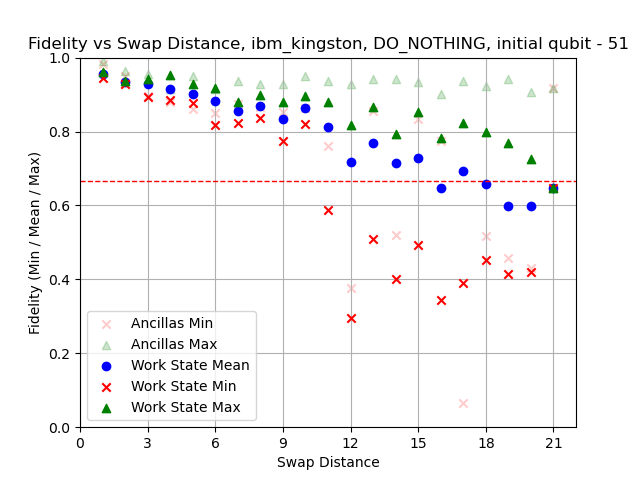}
\caption{IBM Kingston: Initial qubit 51.}
\end{subfigure}
\par\medskip
\begin{subfigure}{0.96\linewidth}
\centering
\includegraphics[width=\linewidth,height=0.34\textheight,keepaspectratio]{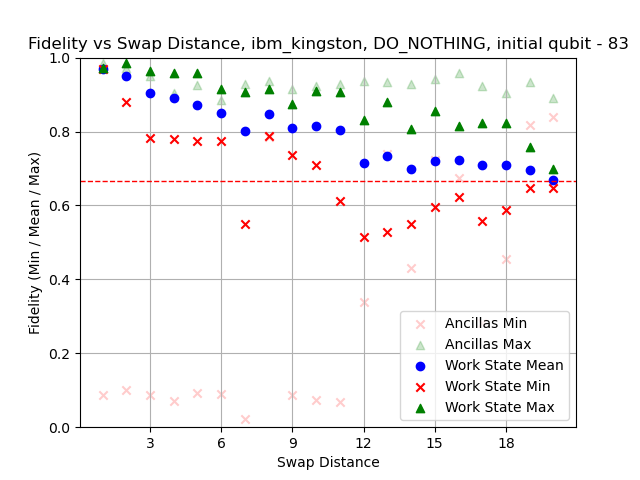}
\caption{IBM Kingston: Initial qubit 83.}
\end{subfigure}
\par\medskip
\caption{Detailed return-score-versus-distance summaries for the initial qubits identified in the panels. In these retained plots, work-state \emph{Fidelity} denotes the measured return score; the $2/3$ line is the empirical cutoff.}\label{fig:detailed_7}
\end{figure}
\clearpage
\begin{figure}[p]
\centering
\begin{subfigure}{0.96\linewidth}
\centering
\includegraphics[width=\linewidth,height=0.34\textheight,keepaspectratio]{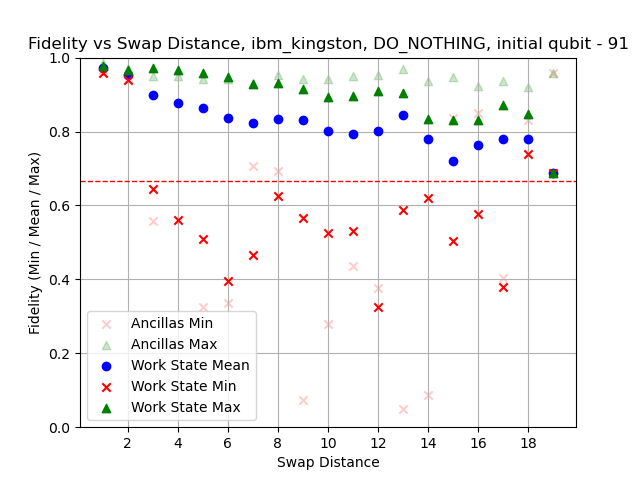}
\caption{IBM Kingston: Initial qubit 91.}
\end{subfigure}
\par\medskip
\begin{subfigure}{0.96\linewidth}
\centering
\includegraphics[width=\linewidth,height=0.34\textheight,keepaspectratio]{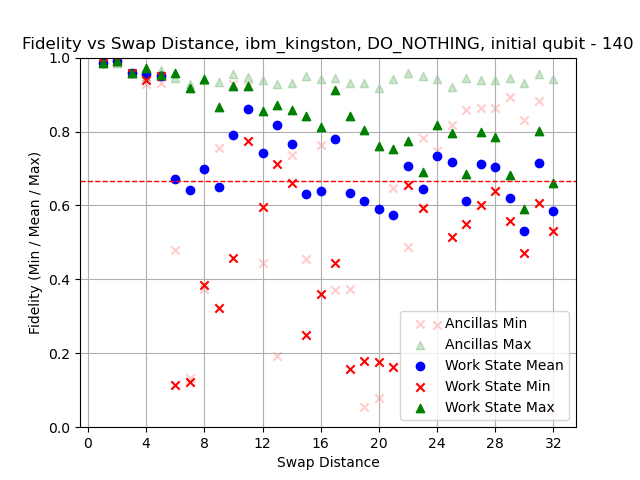}
\caption{IBM Kingston: Initial qubit 140.}
\end{subfigure}
\par\medskip
\caption{Detailed return-score-versus-distance summaries for the initial qubits identified in the panels. In these retained plots, work-state \emph{Fidelity} denotes the measured return score; the $2/3$ line is the empirical cutoff.}\label{fig:detailed_8}
\end{figure}
\clearpage
\begin{figure}[p]
\centering
\begin{subfigure}{0.96\linewidth}
\centering
\includegraphics[width=\linewidth,height=0.34\textheight,keepaspectratio]{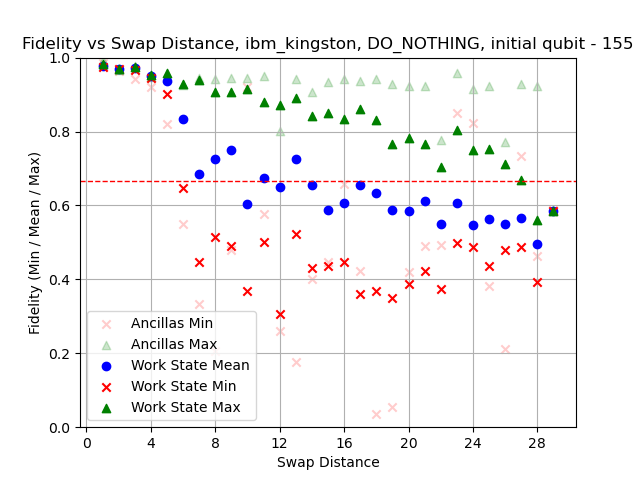}
\caption{IBM Kingston: Initial qubit 155.}
\end{subfigure}
\par\medskip
\begin{subfigure}{0.96\linewidth}
\centering
\includegraphics[width=\linewidth,height=0.34\textheight,keepaspectratio]{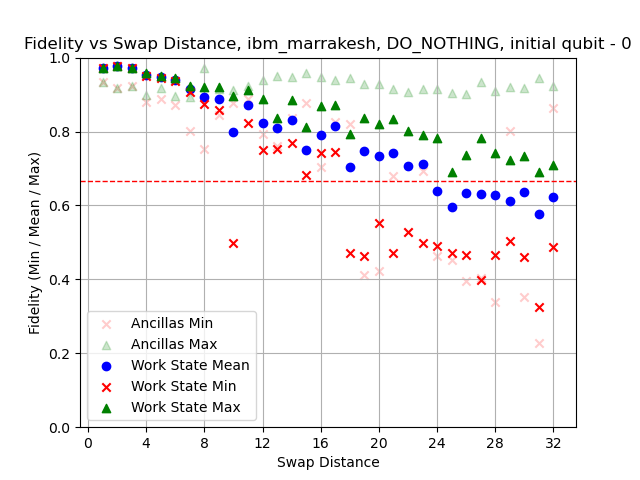}
\caption{IBM Marrakesh: Initial qubit 0.}
\end{subfigure}
\par\medskip
\caption{Detailed return-score-versus-distance summaries for the initial qubits identified in the panels. In these retained plots, work-state \emph{Fidelity} denotes the measured return score; the $2/3$ line is the empirical cutoff.}\label{fig:detailed_9}
\end{figure}
\clearpage
\begin{figure}[p]
\centering
\begin{subfigure}{0.96\linewidth}
\centering
\includegraphics[width=\linewidth,height=0.34\textheight,keepaspectratio]{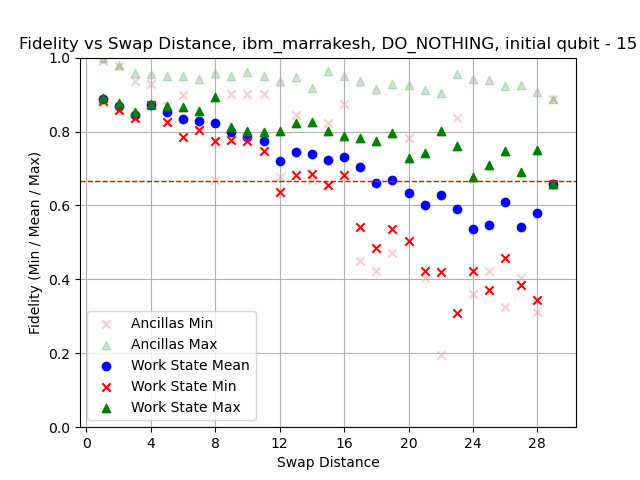}
\caption{IBM Marrakesh: Initial qubit 15.}
\end{subfigure}
\par\medskip
\begin{subfigure}{0.96\linewidth}
\centering
\includegraphics[width=\linewidth,height=0.34\textheight,keepaspectratio]{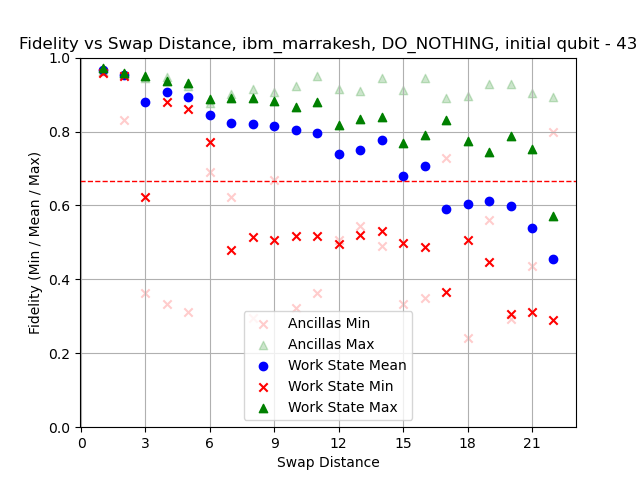}
\caption{IBM Marrakesh: Initial qubit 43.}
\end{subfigure}
\par\medskip
\caption{Detailed return-score-versus-distance summaries for the initial qubits identified in the panels. In these retained plots, work-state \emph{Fidelity} denotes the measured return score; the $2/3$ line is the empirical cutoff.}\label{fig:detailed_10}
\end{figure}
\clearpage
\begin{figure}[p]
\centering
\begin{subfigure}{0.96\linewidth}
\centering
\includegraphics[width=\linewidth,height=0.34\textheight,keepaspectratio]{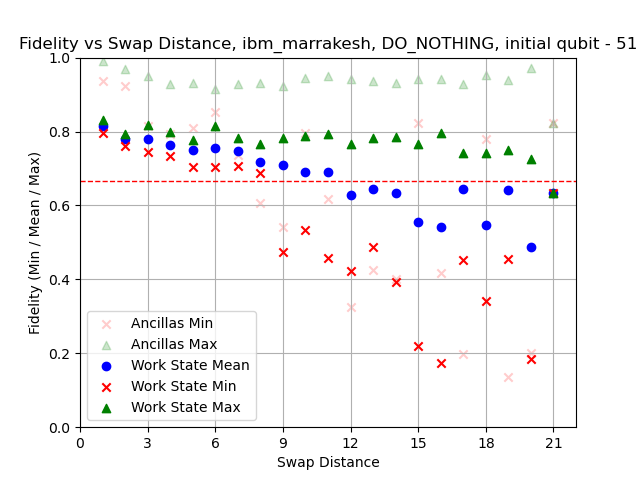}
\caption{IBM Marrakesh: Initial qubit 51.}
\end{subfigure}
\par\medskip
\begin{subfigure}{0.96\linewidth}
\centering
\includegraphics[width=\linewidth,height=0.34\textheight,keepaspectratio]{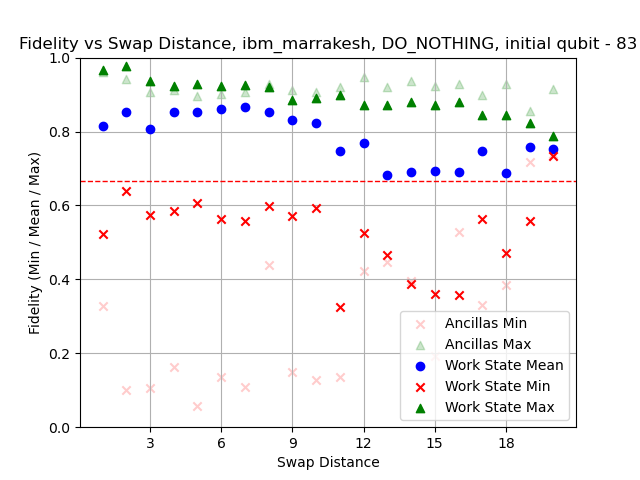}
\caption{IBM Marrakesh: Initial qubit 83.}
\end{subfigure}
\par\medskip
\caption{Detailed return-score-versus-distance summaries for the initial qubits identified in the panels. In these retained plots, work-state \emph{Fidelity} denotes the measured return score; the $2/3$ line is the empirical cutoff.}\label{fig:detailed_11}
\end{figure}
\clearpage
\begin{figure}[p]
\centering
\begin{subfigure}{0.96\linewidth}
\centering
\includegraphics[width=\linewidth,height=0.34\textheight,keepaspectratio]{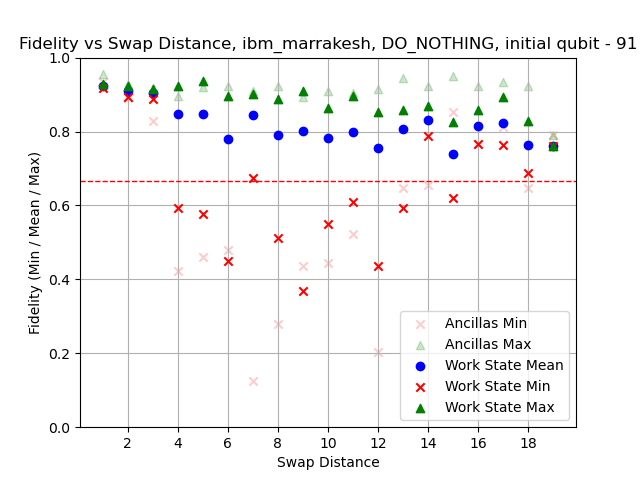}
\caption{IBM Marrakesh: Initial qubit 91.}
\end{subfigure}
\par\medskip
\begin{subfigure}{0.96\linewidth}
\centering
\includegraphics[width=\linewidth,height=0.34\textheight,keepaspectratio]{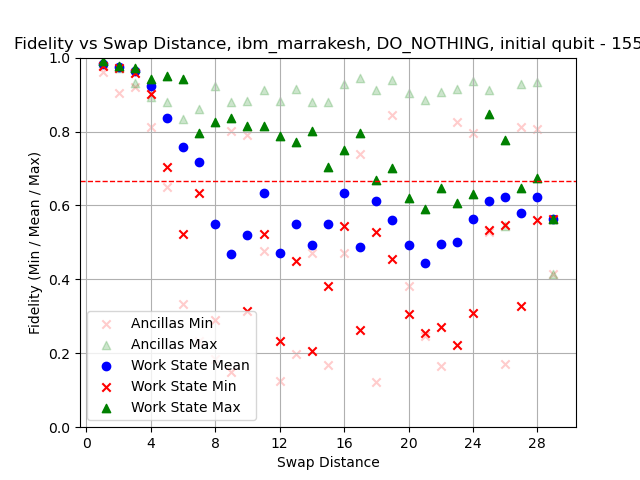}
\caption{IBM Marrakesh: Initial qubit 155.}
\end{subfigure}
\par\medskip
\caption{Detailed return-score-versus-distance summaries for the initial qubits identified in the panels. In these retained plots, work-state \emph{Fidelity} denotes the measured return score; the $2/3$ line is the empirical cutoff.}\label{fig:detailed_12}
\end{figure}
\clearpage

\end{document}